\documentclass{aa}  

\usepackage{graphicx}
\usepackage{txfonts}
\usepackage[dvipsnames]{xcolor}
\usepackage{amsmath}
\usepackage{xspace}
\usepackage{gensymb}
\usepackage{newtxtext}
\usepackage[varvw]{newtxmath}
\usepackage{afterpage}
\usepackage{placeins}

\newcommand{\gam}[0]{$\gamma$\xspace}
\newcommand{\ew}[0]{$\rm EW_{V}/EW_{R}$\xspace}
\newcommand{\rinj}[0]{$R_{\rm inj}$\xspace}
\newcommand{\req}[0]{$R_{\rm eq}$\xspace}

\usepackage{ulem}

\usepackage{hyperref}
\hypersetup{colorlinks,allcolors=black}
\begin{document}

   \title{
   The birth of Be star disks -- \\
   III.  SPH models of localised mass ejections
   }

   \author{A. C. Rubio\inst{1-4},
          A. C. Carciofi\inst{3},
          D. Baade\inst{1},
          J. Labadie-Bartz\inst{5,6},
          T. H. de Amorim\inst{3},
          I. A. Gabitova\inst{3},
          M. W. Suffak\inst{7}
          }

   \institute{European Organisation for Astronomical Research in the Southern Hemisphere (ESO), Karl-Schwarzschild-Str.\ 2, 85748 Garching b.\ M\"unchen, Germany
    \and
    Max-Planck-Institut für Astrophysik, Karl-Schwarzschild-Str. 1, 85748 Garching b.\ M\"unchen, Germany
    \and
    Instituto de Astronomia, Geof{\' i}sica e Ci{\^e}ncias Atmosf{\'e}ricas, Universidade de S{\~ a}o Paulo, Rua do Mat{\~ a}o 1226, Cidade Universit{\' a}ria, 05508-900 S{\~a}o Paulo, SP, Brazil
    \and
    School of Physics and Astronomy, Sir William Henry Bragg Building, Woodhouse Ln., University of Leeds, Leeds LS2 9JT, UK
    \and
    DTU Space, Technical University of Denmark, Elektrovej 327, Kgs., Lyngby 2800, Denmark
    \and
    LIRA, Paris Observatory, CNRS, PSL University, Sorbonne University, Universit\'e Paris Cit\'e, CY Cergy Paris University, 5 place Jules Janssen, 92195 Meudon, France
    \and
    Department of Physics and Astronomy, Western University, London, ON N6A 3K7, Canada
            \\ \email{a.c.rubio@leeds.ac.uk}
             }

   \date{Received XXX; accepted XXX}

 
  \abstract
  {Classical Be stars exhibit mass ejection events that feed their viscous decretion disks. Recent TESS space photometry and simultaneous spectroscopy revealed that these flickers are localised, short-lived, and associated with near-Keplerian rotating material close to the stellar surface.}
  {We aim to constrain the geometrical and dynamical conditions required for a localised surface ejection to generate a Keplerian decretion disk and to predict the corresponding photometric, spectroscopic, and polarimetric observables.}
  {We performed 3D smoothed particle hydrodynamics (SPH) simulations of asymmetric, eruption-like mass ejections from a rotating equatorial sector of a B-type star.  We systematically varied the size and angular velocity of the injection volume, the viscosity of the material, and the injection radius. The SPH outputs were post-processed with the radiative transfer code HDUST to obtain synthetic observables. We scaled the density of each model to approximately match the observed properties of a reference flicker for the Be star f\,Car (B3Vne). }
   {A mildly super-Keplerian rotation of the injection volume, a high viscosity, and a mass-loss rate of the order $10^{-6}\,\rm M_\odot \, yr^{-1} \, str^{-1}$ are required for the ejected material to remain in orbit and form a small Keplerian disk. The resulting photometric, polarimetric, and H$\alpha$ variations reproduce the observed behaviour of the reference flicker. The simulations {confirm} that during mass ejection the disk is strongly asymmetric and dynamically evolving, and circularises within a few days after the end of the flicker, as observed. Models with too wide {mass ejection} angle or too high angular velocity fail to reproduce the observed light curve and line profile behaviour. 
   The models are consistent with mass ejection happening very close to the stellar equator.}
   {Localised, short-duration, mildly super-Keplerian ejections combined with high viscosity and high mass-loss rates can account for the short-timescale variability of the circumstellar environment of Be stars. Be disks {can be} formed from such outbursts and realistic 3D injection geometries are essential to connect surface dynamics to disk build-up within the framework of the viscous decretion disk model. }

\titlerunning{SPH models of localised mass ejections}
\authorrunning{A.C. Rubio, A.C. Carciofi, D. Baade, J. Labadie-Bartz, et. al}
   \maketitle
%

\section{Introduction}

Mass loss happens in all stars throughout their lifetimes, with mass-loss rates generally increasing with stellar mass and luminosity. Even while on the main sequence (MS), O and B-type stars lose a significant amount of mass in the form of radiative line-driven winds, a spherical but inhomogeneous outflow. {For instance, a 35 $\rm M_{\odot}$ star has mass-loss rates on the order of $10^{-6} \rm\, M_{\odot}\, yr^{-1}$ and can shed as much as 40\% of its initial mass during its MS evolution due to winds \citep{renzo2017}.}

In the 1990's, it was suggested that the disks around classical B-emission line (Be) stars originated from such outflows. Classical Be stars are fast rotating MS stars that possess a purely gaseous equatorial disk from where its emission lines originate \citep{rivinius2013}. As these stars had no visible massive companions that could power an accretion disk, Be disks were generally accepted to be outflowing, and so winds seemed to be a suitable cause. However, scenarios where winds were compressed to the equator by the rotational forces of the star \citep{bjorkman1993}, or by wind bi-stability \citep{lamers1991} failed to reproduce the observed Keplerian velocity fields of Be disks\footnote{{We note that these disks cannot be strictly Keplerian, {as they are radially expanding}. Here `Keplerian velocity fields/disks' refer to disks where the azimuthal velocities follow Kepler's third law ($v_\phi \propto r^{1.5}$). }}, and were soon sidelined in favour of the viscous decretion disk (VDD) model \citep{lee1991,bjorkman1997,okazaki2001,bjorkman2005}, in which the disk grows through viscosity. The VDD model makes no assumption on the nature of the process that feeds mass and angular momentum to the disk, other than that it exists.{ Therefore, it cannot be used to directly diagnose the cause of mass loss.} If stellar winds are not responsible for the formation of the disk, another mass-loss process must be at play. This elusive mass-loss mechanism is commonly referred to as the Be phenomenon.

Variability due to circumstellar material is a vital aspect of Be stars: they go through active phases, when they display their characteristic disk-created emission lines, and through quiescent phases, when no emission is detected. Be disks are decretion disks maintained by stellar mass loss; if or when it turns off, the inner disk falls back on the star due to the lack of injection of angular momentum (AM), and the outer disk dissipates. {In this process, there is little mass being truly lost: mass ejected from the star is partially injected into orbit, and partially reaccreted.} Some Be stars, like $\beta$ CMi, have had a stable disk (thus stable mass ejection, at least on average) for decades \citep{klement2015}; others, like $\gamma$ Cas, have oscillated between activity and inactivity several times \citep{baade2023}. 
A portrait of the variability characteristics of Be stars was provided by \citet{figueiredo2025}, who analysed thousands of long-term light curves of Be stars in the Magellanic Clouds.
These episodes of disk build-up and dissipation are particularly interesting given their timescales: a Be disk can grow to dozens of stellar radii in a matter of weeks and dissipate in months. Many Be stars also show shorter episodes of mass ejection that last only hours/days in an outburst-like manner, thus not forming a massive disk \citep{rimulo2018, labadie-bartz2018}; such outbursts are referred to as `flickers'. If indeed all mass ejection events are caused by the same underlying mechanism(s), then it must be able to explain the existence of all these different timescales.

As a class, stars affected by the Be phenomenon have two seemingly omnipresent characteristics: rapid rotation and non-radial pulsations (NRPs). While they are fast rotators, they have been shown not to be critical rotators in their majority \citep[][]{rivinius2006,huang2010}.
The presence of NRPs in early-type Be stars has been known for decades \citep[e.g.][]{baade1982, rivinius1998a, maintz2003, rivinius2003}, and the advent of space photometry has since  strengthened this notion. The recent work of \citet{labadie-bartz2022} presented an analysis of over 400 light curves of Be stars from the Transiting Exoplanet Survey Satellite mission \citep[TESS --][]{ricker2015} and found evidence that all Be stars are non-radial pulsators.

Neither rotation nor NRPs alone can explain the Be phenomenon. Even the combination of these two factors may be insufficient, as there are fast rotating B stars (called Bn stars) in which pulsation is also common, that do not exhibit the Be phenomenon \citep{naze2024} {-- although it is possible that only Be stars pulsate in low-order NRP modes, which could be the key that unlocks the Be phenomenon \citep{penrod1987}.} In any case, it is logical to assume that both pulsation and rotation play a role in the Be phenomenon. Rotation decreases the effective gravity at the stellar equator, making it easier for matter to be lifted into orbit \citep{owocki2006}. Although a single NRP mode has velocity fields limited by the speed of sound (roughly $30 \, \rm km\, s^{-1}$), which by itself is generally insufficient to boost material to orbital velocities, the combination of multiple modes may provide sufficient velocity and AM for launching material. There is observational evidence to support this: in the Be star $\eta$ Cen, \citet{baade2016, baade2018c} found that the outbursts were modulated by a frequency coming from a coupling of modes. This coupling created a mode with a photometric amplitude three times larger than the sum of its parent modes \citep{kurtz2015}, thus able to kick matter into orbit. There are more observational indications of a connection between multi-mode pulsation and mass ejection. For instance, \citet{rivinius1998a} found that the beating of the strongest NRP modes of the Be star $\mu$\,Cen triggers its outbursts; \citet{labadie-bartz2022} detected frequency groups attributed to NRP in the light curves of all active (in terms of exhibiting clear mass ejection episodes) Be stars in their sample.

Establishing rotation and pulsation as the cause of (or correlated to) the Be phenomenon would tie up the apparent ubiquity of NRPs and fast rotation as the necessary (but perhaps insufficient) prerequisites of a Be star. Another frequently proposed mechanism for mass ejection involves small-scale magnetic activity on the stellar surface, capable of lifting material into orbit \citep{nixon2020}. Such fields would be difficult to detect, thus the non-detections of large scale magnetic fields in Be stars \citep{wade2016} do not discard the possibility that they exist and are important for the Be phenomenon. {There is currently no observational data that supports this scenario. In fact, recent observations indicate that material is injected in very low orbits, which would not be the case if it had been lifted by a magnetic lever-arm.}

Neither of the two main physical scenarios for the Be phenomenon---magnetic fields or NRPs---has yet reached a level of maturity that allows for detailed modelling of the interface between the outer layers of the star and the disk. A recent observational effort attempted to bring the problem into focus from an empirical standpoint. \citet{labadie-bartz2025} investigated short-lived `flickers' in 13 Be stars observed simultaneously with TESS light curves and time-series spectroscopy, to which six stars from the literature were added. This dataset allows, for the first time, a systematic investigation of the Be phenomenon over the key timescales of rotation, pulsation, and near-surface orbits, spanning approximately 0.5 -- 3\,d.

These newly acquired data enables us to explore a simple, yet realistic, mass ejection scenario: an isotropic mass ejection from a rotating injection region on the surface of the star \citep[first proposed by][but not developed further]{kroll1997}. Using smoothed particle hydrodynamics (SPH), we simulate these outbursts in 3D, exploring several parameters for the angular frequency of the ejection site and the isotropic velocity of the ejected particles. We qualitatively compare the expected light curves, H$\alpha$ profiles and polarimetric signals from these simulations to the data, marking the first time SPH simulations of Be star mass ejections are confronted with well-documented events. Our goal is to explore what are the necessary mass ejection geometry and dynamics to build a Be disk, assuming mass is lost from a region in the equator of the star. We provide an overview of the observational trends expected from these events, thereby laying a foundation for future investigations into the physical origin of the Be phenomenon.

The paper is structured as follows: in Section~\ref{sec:review} we present an overview of the observational findings of \citet{labadie-bartz2025}; in Section~\ref{sec:model_descrip} we describe our modelling procedure, followed by our results in Section~\ref{sec:results}. Finally, our discussion and conclusions are in Section~\ref{sec:discussion}.

\section{Review of observational signatures}\label{sec:review}

\citet[][hereafter Paper I]{labadie-bartz2025} present a detailed description of the observational behaviour of outbursting Be stars, with simultaneous TESS space photometry and spectroscopy from the Network of Robotic Echelle Spectrographs \citep[NRES --][]{brown2013}, Dominion Astrophysical Observatory \citep[DAO --][]{monin2014}, CHIRON \citep{tokovinin2013} and Be Star Spectra database \citep[BeSS,][]{neiner2011}. The full sample comprises 13 Be stars. The targets were selected based on their brightness, disk status, and history of activity, to increase the probability of observing a full {mass ejection} event with TESS. As the TESS data are not immediately made available after observation, the ground-based follow-up had to be made blindly, without prior knowledge of ongoing events.

The observed events were coarsely divided into three distinct categories (`clear', `pristine', and `complex') depending on how well-defined the outbursts were and whether or not there was a pre-existing disk at the time.
The best examples of each category, V767\,Cen, f\,Car, and 12\,Vul, were described in detail in Paper~I: V767\,Cen already had a disk at the time a well-defined, `clear' new outburst began, f\,Car is a `pristine' case having no pre-existing disk at the time of the observed outbursts, and 12\,Vul had a pre-existing disk and quasi-continuous and `complex' mass ejection such that individual events could not be separated. Both V767\,Cen and f\,Car were observed in two consecutive TESS sectors, providing a baseline of about 50 days, and their spectroscopic campaigns started a few days before the beginning of the TESS run and continued for a few days afterwards. 12\,Vul was only observed simultaneously by TESS and spectroscopy for one sector (about 27 days). 

The most relevant observational signals found in the data are the amplitudes and timescales of the photometric response to mass ejection, the overall strength of the emission lines (particularly H$\alpha$), measured by their equivalent width (EW), the variations in the relative EW of the violet and red sides of the line (\ew), and the distance between the two emission peaks (peak separation, PS). Following Paper I, \ew is defined as
\begin{equation}\label{eq:ew}
    \mathrm{\frac{EW_V}{EW_R}} = \frac{\int^{\lambda_0}_{\lambda_1} (2 - F_\lambda)d\lambda}{\int^{\lambda_2}_{\lambda_0} (2 - F_\lambda)d\lambda} \,,
\end{equation}

\noindent where $\lambda_1$ and $\lambda_2$ are the outer integration limits (in wavelength), $\lambda_0$ is the central wavelength of the line correct for the systemic velocity of the star ($v_0$), and $F_\lambda$ is the normalised flux.

The PS is another classic and fundamental observable for understanding the dynamics of the circumstellar disk in Be stars. For instance, it can be used to estimate the inclination angle of the disk and the size of its emitting area \citep[e.g.][]{sigut2023, hanuschik1986}.
Similarly to the line asymmetry, instead of the traditional PS, we use an alternate definition, proposed by \citet[][Eq. 2]{carciofi2025}:
\begin{equation}\label{eq:psw}
 \mathrm{PS_W} = \frac{\int_{v_0}^{w+v_0} v {F_v} dv}{\int_{v_0}^{w+v_0}  {F_v} dv} - \frac{\int_{-w-v_0}^{v_0} v {F_v} dv}{\int_{-w-v_0}^{v_0}  {F_v} dv}\,,
\end{equation}
\noindent where $w$ is an integration limit in velocity and $F_v$ is the normalised flux. This equation measures the separation between the velocity-weighted flux averages, so $\rm PS_W$ can be interpreted as the separation in velocity between the photocenter of each side of the line.  
{Because our focus is on the disk dynamics, we restrict the analysis to the velocity interval, $[-w, w]$, where most of the relevant emission and disk absorption features are expected to occur: we thus adopt for $w$ the projected orbital velocity, $v_{\mathrm{orb}} \sin(i)$, with $v_{\rm orb} = (GM/R_{\rm eq})^{1/2}$, which is $\sim 500 \rm \, km \, s^{-1}$ for the central star we use in this work.}

\begin{figure*}
    \centering
    \resizebox{\hsize}{!}{\includegraphics{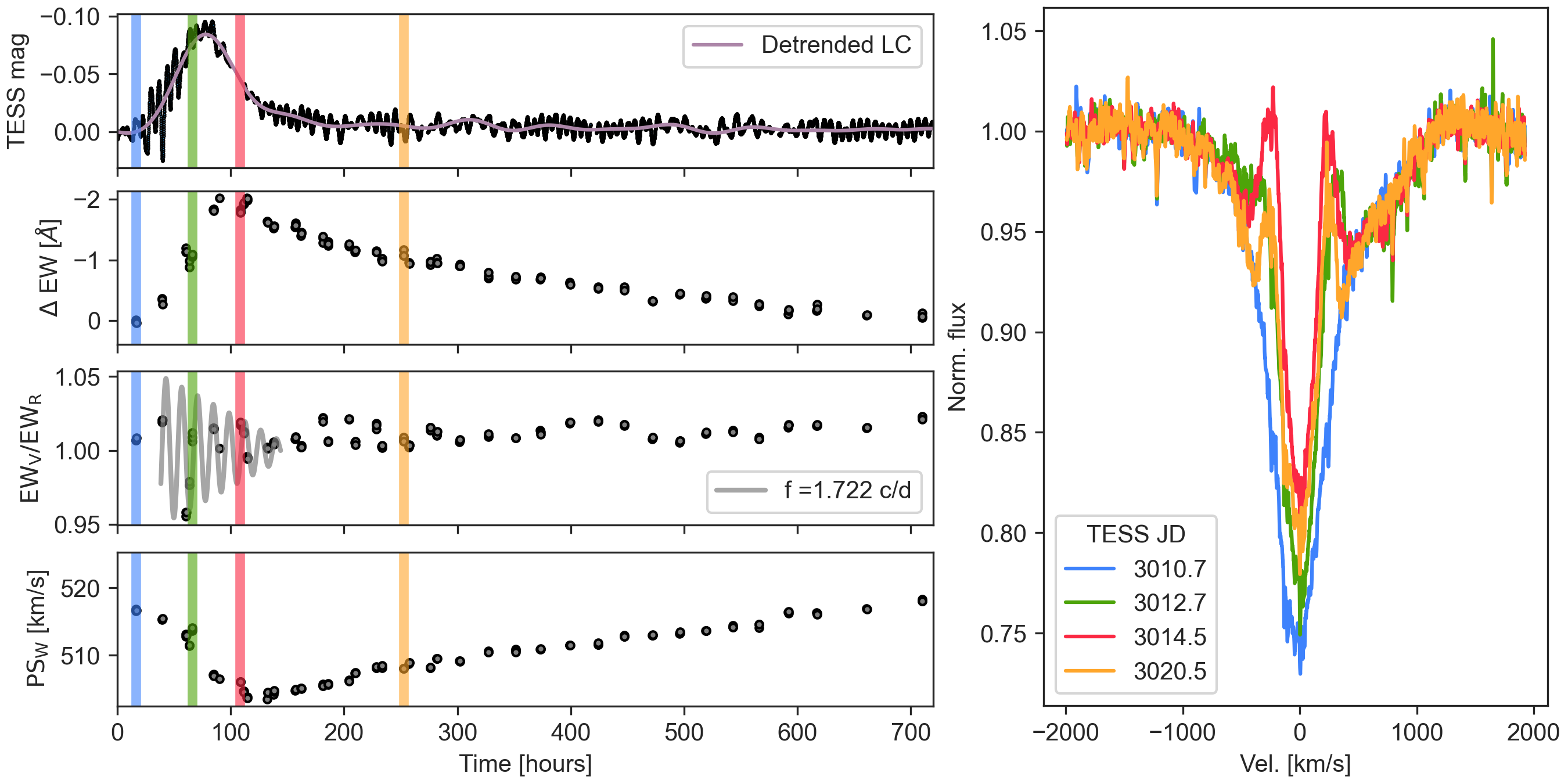}}
    \caption{Observational data for the Be star f\,Car, from Paper I. The top-left panel: TESS light curve, bottom-left panels show EW, \ew and $\rm PS_W$ for the H$\alpha$ profiles obtained with NRES. 
    {In the top panel, the grey line is the light curve with high-frequency variability (> 0.5 $\rm d^{-1}$) removed.}
    In the third panel, the grey line is the fit of the \ew to a sinusoidal function. The right panel: H$\alpha$ profiles at four epochs, marked in the left panels by the coloured bands in the same colours. 
    }
    \label{fig:fcar}
\end{figure*}

Although V767\,Cen, f\,Car, 12\,Vul, and the rest of the sample differ in the pre-existence of a disk and the number, strength, and duration of outbursts, the following characteristics are consistently present for all targets\footnote{The observational characteristics of Be+disk systems are inclination dependent. As the targets analysed in Paper~I are not seen edge-on (inferred, as H$\alpha$ emission is always accompanied by a brightening, not a fading, in their light curves), the list we provide is biased towards low inclination systems. 
The effects of inclination on the observables are explored in Sect.~\ref{sec:results}.}:

\begin{enumerate}
    \item {Pulsations are detected in photometry organised in one or more frequency groups; }
    \item The morphology of the photometric flicker follows a consistent pattern: a rapid rise during outburst, followed by a slower decrease and eventual return to the baseline (i.e. the pre-flicker brightness). An example is shown in Fig.~\ref{fig:fcar};
    \item Emission in the H$\alpha$ line profile takes longer to disappear than the photometric signal;
    \item At the onset of an event, changes in photometry and spectroscopy are nearly concurrent; 
    \item Emission line asymmetries (variations in \ew) are present, arising with photometric variation and evolving rapidly throughout the outburst. The variability of these asymmetries is cyclic;
    \item For the same star, the frequency of these asymmetry oscillation cycles remains nearly constant across distinct mass ejection events, even if the events themselves are longer or stronger;
    \item As far as could be determined, the \ew asymmetry frequencies do not change during a given event even as they dampen out after mass injection stops (or greatly decreases);
    \item The damping of the \ew asymmetries happens during the decreasing phase of the flicker, usually after 5 -- 10 cycles;
    \item The emission oscillation frequencies are always around 10\%--20\% lower than the main spectroscopic pulsational frequency of the star;
    \item The \ew frequencies are consistent with the orbital frequency near the stellar equator, indicating that the material responsible for this asymmetry is in orbit but still quite close to the equator.
    
\end{enumerate}

\noindent 
The combined photometric and spectroscopic evidence indicates that flickers are localised, discrete, short-lived mass-ejection episodes that feed the disk, rather than large-scale or stochastic processes. Additionally, the repeatability of frequency patterns across different events suggests the action of a stable underlying mechanism, likely linked to near-surface stellar dynamics.
Taken together, these observational characteristics summarily rule out mass ejection scenarios that do not generate asymmetrically distributed material in orbit close to the stellar equator.

Although all targets in the sample exhibit the ten characteristics described above (considering their isolated outbursts), we have selected f\,Car {(HD\,75311, HR\,3498)} as the prime example of the observational effects of an asymmetric mass ejection. This choice is motivated by the fact that it underwent two distinct short outbursts and showed no evidence of a pre-existing disk. In our analysis, we focus on the stronger of the two, the second outburst.

The observational data from TESS and line measurements from various NRES spectra are shown in Fig.~\ref{fig:fcar}.
In the hours following the start of the flicker, the matter is highly azimuthally concentrated and rotates around the star, as evidenced by strong variations in the \ew of the H$\alpha$ profile. After $\sim7$ days, the variations dampen as the matter spreads and circularises. $\mathrm{PS_W}$ also displays strong variations during the flicker, rising and decreasing, {mirroring the behaviour of the EW}.

Paper II in this series investigated the behaviour of one of the stars introduced in Paper I, $\lambda$\,Pav, but in greater detail with an emphasis on the observed photospheric and circumstellar variability before, during, and after an outburst \citep{nova2026}. Despite the excellent spectroscopic coverage, $\lambda$\,Pav was not an ideal point of comparison for the modelling work presented here because only the first few days of the outburst were sampled with photometry by TESS.

\section{Model description}\label{sec:model_descrip}

Most previous studies of mass ejection events in Be stars have treated the problem in one dimension \citep[e.g.][]{rimulo2018, ghoreyshi2018}.
Even intrinsically 3D simulations (such as the SPH studies of \citet{okazaki2002} and many others that followed using the same code) employ a simple geometry for the mass ejection in which mass is injected into the disk in a ring around the star. 
With these simple mass ejection assumptions, these works have had success in reproducing the overall large scale behaviour of Be disks \citep[e.g. ][]{suffak2020, rubio2025}. The results of Paper I, however, indicate that mass ejection is azimuthally asymmetric, at least for Be stars in the sample, selected for their history of activity. The flickers studied in Papers I and II were also short duration, lasting from a few hours to days{, whereas previous modelling efforts focus on months to decades long outbursts}. Therefore, a new tool is needed to tackle this problem in a more realistic way, one that can account for the complex geometry and temporal variability of these ejection events.

We modified the SPH code of \citet{okazaki2002} to allow for short-lived, asymmetric, eruption-like mass ejection that happens over a section of the equator. 
Once the material reaches an orbit, it evolves under viscous interactions. The viscosity ($\nu$) is parametrised according to the $\alpha$-disk formulation of \citet{shakura1973}: $\nu = \alpha c_s H$, where $c_s$ is the sound speed in the disk, $H$ is the disk scale height, and $\alpha$ is the viscosity parameter, ranging from 0 to 1, {invariable with time and throughout the disk.} We also assumed that the injected material remains isothermal throughout the simulation. {We highlight that, differently from works that used the original version of the code, we do not include a binary companion in any of our simulations.}

\begin{figure}
    \centering
    \includegraphics[width=.8\linewidth]{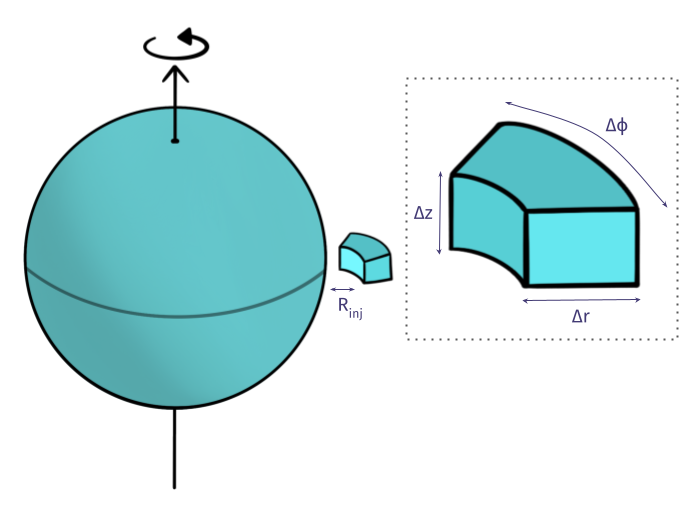}
    \caption{Schematic of the injection zone. The parameters  $\Delta z$, $\Delta \phi$, $\Delta r$, and $\rm R_{inj}$ are defined. This sketch is not to scale. }
    \label{fig:esquema}
\end{figure}

The new parameters added to the code describe the geometry and kinematics of the injection volume: in cylindrical coordinates, these include the radial extent ($\Delta r$), vertical height ($\Delta z$, measured from the equatorial plane, up and downwards), and azimuthal opening angle ($\Delta \phi$, {a range in azimuth}). Additionally, the injection radius ($R_{\rm inj}$) defines the centre of the injection volume, $v_i$ specifies the isotropic ballistic speed of the particles, and $\gamma$ sets the rotational speed of the injection volume.
This injection volume is, therefore, not anchored to the stellar surface, and can rotate independently of the stellar surface. Our injection volume is simply a construct to insert particles in the simulation, as we make no assumptions on how the star is flinging the material into orbit. A schematic of the injection volume is in Fig.~\ref{fig:esquema}.
We define \gam as a multiplicative factor of the orbital speed at the injection radius, so that the angular frequency of the injection volume and the particles is
\begin{equation}
\Omega_{\rm inj} = \gamma \, \Omega, 
\end{equation}
where $\Omega = (GM/R_{\rm inj}^3)^{1/2}$. The injection radius \rinj can have any value larger than $R_{\rm eq} + \Delta r/2$, where \req is the equatorial radius of the star.

In our formulation, all particles appear in a random position inside the injection volume with a velocity that comprises a rotational component based on \gam, plus an isotropic ballistic component, $\bf{v}_i$, whose {modulus} is the same for all particles. The direction of $\bf{v}_i$ is randomised over a hemisphere oriented away from the stellar surface (as defined in App.~\ref{apd:def}) rather than over a full sphere.

The rationale behind our choice of parameters is based on the well-established physics of Be disks (following the VDD model) and the observational behaviour seen in Papers I and II. The timescale of the spread and growth of the ejected material from a localised overdensity into a ring, and then a disk, is governed by viscosity, the main force acting on the material other than the gravity of the Be star. Thus, exploring different values of the viscosity parameter $\alpha$ changes the timescale of the evolution of the observables. The size and location of the injection volume set the initial geometry of the ejecta before viscous spreading drives the latter towards symmetry.
Since this initial asymmetry shapes the variations in \ew, and the ejecta velocity influences $\rm PS_W$, our goal was to constrain the maximum azimuthal extent of the injection volume and to investigate how the observables respond to changes in the radial and vertical dimensions of the injection volume.

The total velocity of each particle is given by the vector sum of its orbital component and $\bf{v}_i$. Therefore, a particle may or may not have enough angular momentum to be lifted into orbit. Ultimately, this initial velocity together with the effects of viscosity determines how much material actually stays in orbit and how much falls back into the star in the following time step. 
If not enough material survives in orbit, the disk has too low a density {and emitting area} to account for the brightness variation and line emission seen in the observations presented in Papers I and II. As such, we aimed to constrain the velocity configuration necessary for the formation of a substantial disk, in line with the observations.

\begin{table}[]
    \centering
    \caption{SPH simulation parameters.}
    \begin{tabular}{c|c}
    \hline
        Variable parameters & Values \\
        \hline
        \hline
        $\alpha$ & 0.1 and 1.0*\\
        \gam  & 0.9, 1.0, 1.05*, 1.2\\
        $\Delta \phi$ & 0.2*, 2.0, 4.0, 6.0 rad\\
        $\Delta z$ & 0.2*, 1.0 $R_{\rm eq}$\\
        $R_{\rm inj}$ & 1.01*, 1.25 $R_{\rm eq}$\\
        $v_{i}$ & 20*, 50 $\rm km \, s^{-1}$\\
        \hline
        Fixed parameters & \\
        \hline 
        \hline
        $\Delta r$ & 0.01 $R_{\rm eq}$\\
        $M$ & 7.6 $\rm M_{\odot}$ \\
        $R_{\rm eq}$ & 5.8 $\rm R_{\odot}$ \\
        $T_{\rm eff}$ & 20 kK \\
        $v_{\rm orb}$ & 499.7 $\rm km \, s^{-1}$\\
        $\dot{M}_{\rm ejec}$ & $10^{-7} \, \rm M_\odot \, yr^{-1}$\\
        $t_{\rm outburst}$ & 42.4 h\\
        \hline
    \end{tabular}
    \tablefoot{The parameters {values} marked with `*' are the ones of our preferred model, explained in detail in Section~\ref{sec:most_fave}. }
    \label{tab:params}
\end{table}

We ran a suite of simulations exploring the parameters detailed in Table \ref{tab:params}. {We used $M = 7.6\,\rm M_{\odot}$ and $R = 5.8\,\rm R_{\odot}$ for the sink particle that represents the central star in the SPH simulations, and a disk temperature of $T_{\rm d} = 0.6 \,T_{\rm eff} = 12\,\rm kK$. These values are based on the estimated parameters derived by \citet{zorec2016} for f\,Car. All simulations were initialised with no circumstellar material; mass ejection was turned on instantaneously and was active for 42.4\,h (three complete orbits of the star).}

The density and velocity distributions from the SPH simulations were mapped onto a 3D grid and given as input to the radiative transfer code HDUST \citep{carciofi2006, carciofi2008}, following \citet{suffak2024}. HDUST solves the thermal and statistical equilibrium conditions in the NLTE regime, thereby providing the electron temperature and the hydrogen level populations across the grid. Using these results, HDUST computes the emergent spectrum and produces synthetic photometric and polarimetric signals, H$\alpha$ line profiles, and images.

\section{Results}\label{sec:results}

We first present an overview of the model that best matched the observational data of the reference flicker of f\,Car, detailing the SPH simulation results and the radiative transfer calculated observables. This model was selected by visual comparison between data and all models calculated for this work. In the following subsections we describe the effects of changing key parameters of the mass ejection on these observables: the rotational velocity of the injected particles, the opening angle {and vertical extent} of the injection volume, the disk viscosity, and the injection radius. We note that each simulation (SPH and radiative transfer) is time and computationally expensive, given their inherent complexity. Therefore, we focused on exploring the sensitivity of the observables to the input parameters of the model, and achieving a qualitative representation of the observables rather than performing a full quantitative model-fitting.

\subsection{Overview of the most favoured scenario}\label{sec:most_fave}

The model that best reproduced the general behaviour seen in the reference flicker of f\,Car (Sect.~\ref{sec:review}, Fig.~\ref{fig:fcar}) has \gam = 1.05, $\alpha = 1.0$, and $R_{\rm inj}$ = 1.01 \req, i.e. very close to the stellar equator, with $\Delta r = 0.01$ \req, $\Delta z = 0.2$ \req, and $\Delta \phi = 0.2$ rad.
These parameters {values} are indicated by the `*' symbol in Table~\ref{tab:params}.
In this subsection, we present a detailed description of the structure and dynamics of the circumstellar environment, followed by an overview of the HDUST predictions.

\subsubsection{Structure and dynamics}

\begin{figure*}[h!]
    \centering
    \resizebox{\hsize}{!}{\includegraphics{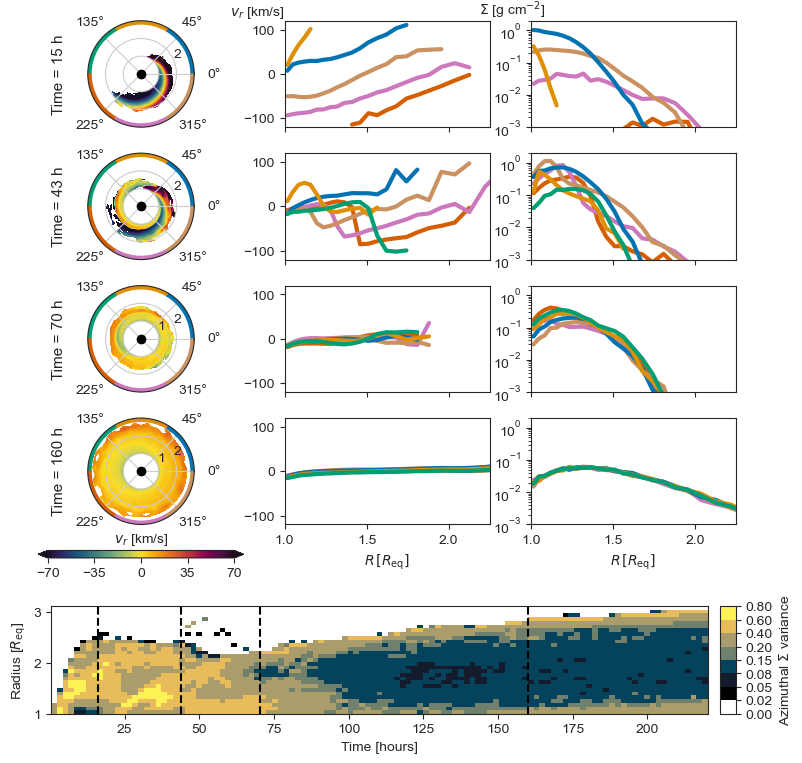}}
    \caption{Evolution of the simulation with \gam = 1.05, $\alpha = 1.0$, $R_{\rm inj}$ = 1.01, $\Delta r = 0.01$ \req, $\Delta z = 0.2$ \req, and $\Delta \phi = 0.2$ rad. The first four rows show the 2D radial velocity map (polar plots), and line plots for the radial velocity and surface density ($\Sigma$) for four distinct snapshots (top to bottom, 15, 43, 70 and 160\,h). Six wedges (each 60\degree wide) are defined in the polar plots by the colours on their borders (these coloured borders are in the observer's frame of reference, i.e. they are not corotating with the star). The line plots show the average radial velocity ($v_r$, the velocity with which particles move away from the central star, radially) and surface density ($\Sigma$) contained in each of these wedges, to illustrate the evolution of the azimuthal distribution of material around the star. The bottom plot is a map of the temporal evolution of the azimuthal variance (over 360\degree) of $\Sigma$ (in percent) with time: the lower the variance, the `smoother' and more even the disk is. 
    {In broad terms, lighter colours indicate a more azimuthally inhomogeneous disk, while darker colours correspond to a smoother azimuthal distribution.}
    }
    \label{fig:bestsph}
\end{figure*}

\begin{figure}
    \centering
    \resizebox{\hsize}{!}{\includegraphics{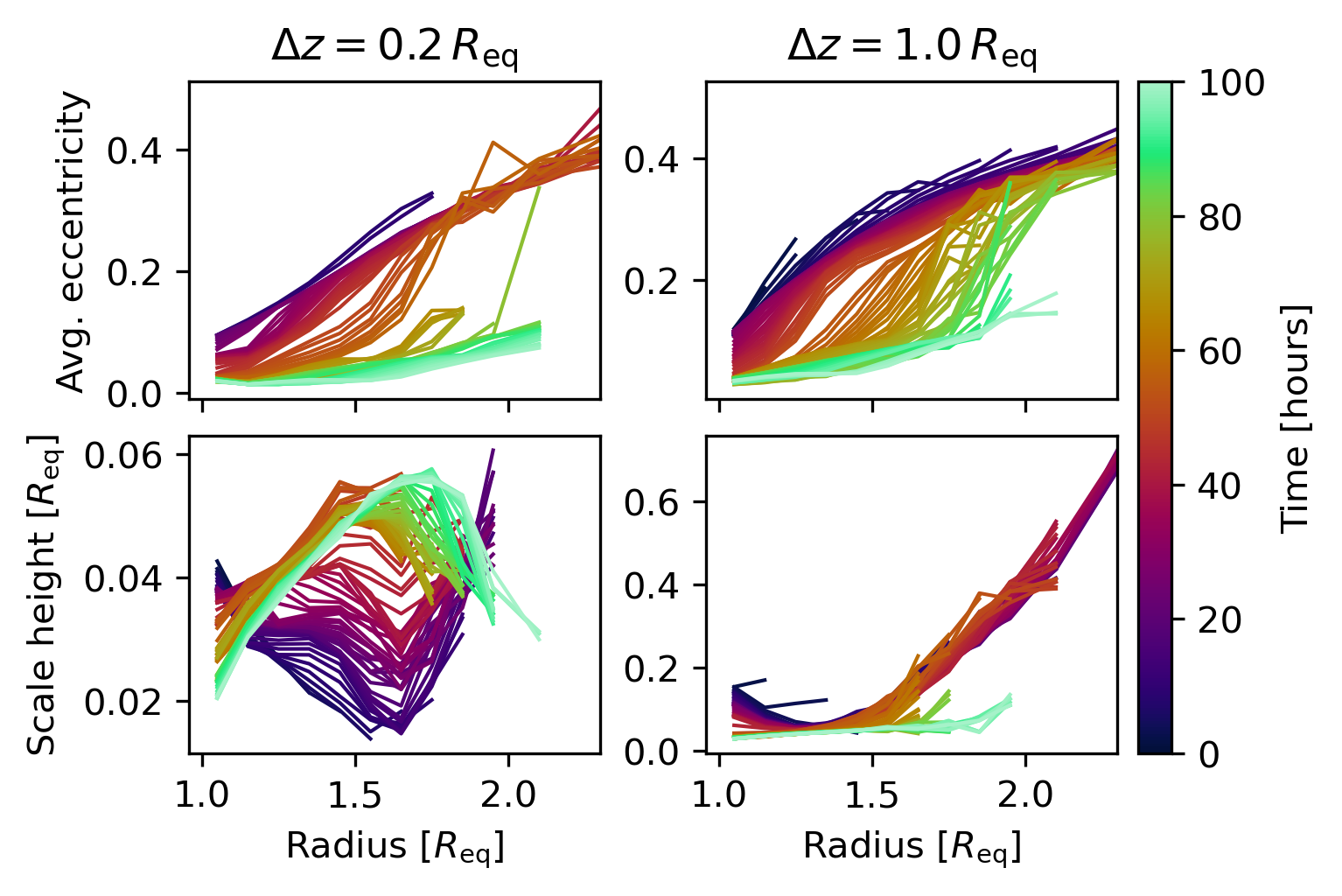}}
    \caption{Average values of eccentricity of the particles orbits (top panel) and disk scale height (bottom panel) inside radial bins, evolving with time, as denoted by different colours. The first column shows the model described in Sec.~\ref{sec:most_fave}, and the second shows the model with larger $\Delta z$, discussed in Sec.~\ref{sec:height}. The eccentricity drops with time while the scale height goes from a higher inner disk to a flaring disk after mass ejection stops.}
    \label{fig:eccs}
\end{figure}

Our Fig.~\ref{fig:bestsph} presents an overview of the evolution of the simulation from the beginning of mass ejection ($t = 0\, \rm h$) to the formation of a small Keplerian disk around the star. During mass ejection, and persisting for tens of hours after its termination, the orbits of the particles are not circular (Fig.~\ref{fig:eccs}), leading to these asymmetries in density. Two processes collaborate to circularise the material into a Keplerian disk: orbital phase mixing and viscosity. Orbital phase mixing implies that particles with slightly different angular velocities (due to varying radial distances or random velocity components) drift apart in azimuth over time, thus smoothing out azimuthal inhomogeneities. Meanwhile, the internal stresses in the disk (modelled as viscosity) redistribute AM, allowing particles on eccentric orbits to settle into more circular, stable orbits. This viscous diffusion happens in a timescale that follows \citep{bjorkman2005}
\begin{equation}
    t_{\rm diff} = \frac{r^2}{\nu} = \frac{r^2}{\alpha c_s H} \sim \rm days,
    \label{eq:visctimescale}
\end{equation}
\noindent where $r$ is the radius of the orbit and $H$ is the typical disk height at that location. Together, orbital phase mixing and viscosity act to transform the initial, irregular mass distribution into a more azimuthally symmetric, circularly rotating disk that slowly expands outwards (see bottom panel of Fig.~\ref{fig:bestsph}). In other words, azimuthal homogenisation of the material immediately follows from the circularisation of the orbits of the particles. When we refer to `circularisation', below, we are discussing the combination of these effects.

In the first snapshot (top row of Fig.~\ref{fig:bestsph}), the Be star has been undergoing continuous mass ejection for 15\,h. During that time span, the injection volume completed 1.1 orbits, corresponding to a rotation of 430$\degree$ around the star.
All particles are injected with a velocity given by Eq.~\ref{eq:direc2}, with $|v_{i}| = 20\,\rm km\,s^{-1}$ ({roughly the sound speed in the disk}), and have a random direction pointing away from the surface of the star, as defined in Sect.~\ref{sec:model_descrip}. 
Thus, every particle has enough velocity, initially, to stay in orbit because the smallest azimuthal velocity, $\gamma \, v_{\rm orb}-|v_{i}| = 504.7 \, \rm km \, s^{-1}$, is larger than $v_{\rm orb} = 499.7 \, \rm km \, s^{-1}$.
The continuous input of AM {into} the injection volume pushes part of the material outwards. The particles in more external orbits gain AM through shear interactions with the particles in inner orbits. The net result is the formation of an expanding front with high radial velocities and an infalling region with negative radial velocities. Both structures trail behind the injection volume (top left panel), illustrating the dynamical response of the disk to the localised mass ejection.

The second snapshot (second row of Fig.~\ref{fig:bestsph}) corresponds to the moment when mass ejection has just terminated. At this point in the simulation, there is material around the entire equator of the star. Its density, however, still has a strong azimuthal dependence, with material mostly concentrated around the injection volume. As in the first snapshot, the mass injection spiral is anchored on the injection volume (located at 45$\degree$ in this snapshot), which has already completed 3 orbits. We see that a second spiral arm of positive $v_r$ (yellow in the $v_r$ map) has formed, trailing the main spiral. This behaviour indicates that the dynamics of the mass decretion and re-accretion is not as simple as a gravity filter where all material with negative $v_r$ falls back into the star. A possible explanation for this more complex behaviour is that the infalling material ($v_r < 0$) still has significant specific AM. 
As it drops in orbit, this AM-rich material interacts with the particles already present at the lower orbits. The AM is redistributed among the particles via shear viscosity. Therefore, although many particles do fall back to the star, some acquire enough AM to stay in orbit, leading  to the accumulation of material with low, but positive $v_r$ in low orbits. This accumulation trails behind the injection volume by approximately 45$\degree$.

At 70\,h (third row of Fig.~\ref{fig:bestsph}), no new mass has entered the disk for 30\,h. Without the constant input of mass and AM, the injection volume stops being the anchor point of the forming disk. The material reorganises itself by the viscous interaction between layers, and the radial velocity begins to settle into values of a few kilometres per second in all azimuthal directions. The density also begins to converge to a similar profile in all directions. At this point, the material approaches circularisation, where the memory of the initial position, density, and velocity of the injected material is lost.

Nearly 5 days after the end of mass injection, at 160\,h, both the radial velocity and the density have been smoothed around the star, forming a uniform disk (fourth row of Fig.~\ref{fig:bestsph}). The radial extent of the disk has increased by $\approx 1.0 \, \rm R_{\rm eq}$ since the previous snapshot, as the material travelled from lower to higher orbits. 
Most of the disk is now azimuthally symmetric, with radial variance between 5 and 15$\%$ (see the variance map in Fig.~\ref{fig:bestsph}). The exception is the outermost region of the disk, where the azimuthal density fluctuations are still significant. We propose that this effect arises from a ‘circularisation front’ propagating outward while the regions not yet reached by this front remain more inhomogeneous. {This effect can be seen in the eccentricity and scale height evolution plots of this model, in the leftmost panels of Fig.~\ref{fig:eccs}}.

\subsubsection{Deviations from standard viscous decretion disk theory}

The current most used theory for the disks of Be stars is the VDD model. In its isothermal, steady-state solution \citep[][]{bjorkman2005}, the disk is considered to be 1) in hydrostatic equilibrium (no vertical motion, $v_z = 0$), 2) thin ($H << r$), and 3) composed of material on circular orbits around the central Be star. 
In our simulations, none of these conditions apply until the orbiting material relaxes, which happens \~60\,h after the end of mass injection.

The mass injection events we simulate are asymmetric and ballistic in nature, launching material not only in the tangential direction but also with substantial vertical components, both above and below the equatorial plane. A key consequence is that the circularisation and vertical hydrostatic equilibrium timescales are comparable. The former is on the order of days, as seen above. The latter can be estimated as the time a {sound wave} takes to travel through the disk vertically, i.e.
\begin{equation}
\tau_{H} \sim \frac{H}{c_s} \,.
\end{equation}
$\tau_{H}$ is on the order of hours, especially in the inner disk. As a result, the vertical structure cannot be assumed to settle instantaneously. Indeed, the scale height of the material close to the star falls as the disk evolves and begins to flare (see the lower-left panel of Fig.~\ref{fig:eccs}). 
Hence, it is not appropriate to describe the inner disk as being in vertical hydrostatic equilibrium, nor to assume that it is thin near the star, as is commonly done in steady-state VDD models.

Furthermore, the orbits of the particles during the mass injection phase — and even for a significant period afterward — remain non-circular, as indicated by the radial velocities in Fig.~\ref{fig:bestsph} and by the eccentricities in Fig.~\ref{fig:eccs}. The combination of incomplete circularisation, vertical motion, and azimuthal asymmetries means that the inner disk during active mass injection is dynamically evolving and far from the idealised state represented by standard VDD theory.

Taken together, these results indicate that during its active phase, particularly near the star, the disk of a Be star is not well described by the assumptions of the steady-state VDD framework. Instead, a full treatment of the injection geometry, 3D kinematics, and time-dependent evolution is essential to accurately capture the physical state of the system.

\subsubsection{HDUST observables}

\begin{figure*}
    \centering
    \resizebox{\hsize}{!}{\includegraphics{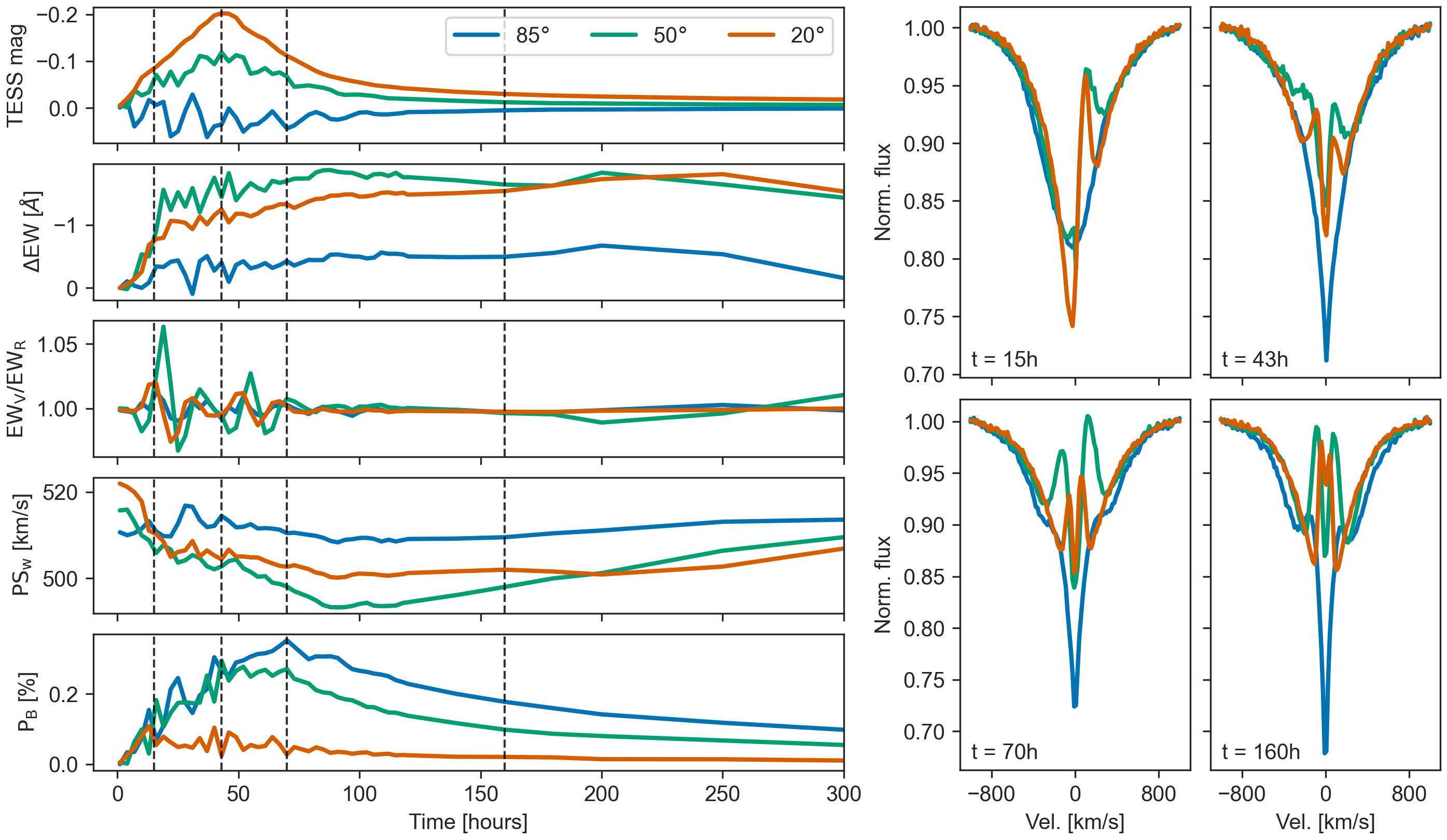}}
    \caption{Photometry and H$\alpha$ line measurements for our preferred model with $\gamma = 1.05$, $\alpha =1.0$, $R_{\rm inj}$ = 1.01, $\Delta r = 0.01$ \req, $\Delta z = 0.2$ \req, and $\Delta \phi = 0.2$ rad. Colours indicate different inclination angles. Panels on the right show the H$\alpha$ profile for the four snapshots highlighted in Fig.~\ref{fig:bestsph}: 15\,h, 43\,h, 70\,h and 160\,h, also marked in the left-hand panels by the dashed vertical lines.}
    \label{fig:besthdust}
\end{figure*}

When a Be star becomes active, the ejected material manifests observationally in photometry, spectroscopy, and polarimetry as it scatters, absorbs, and re-emits stellar flux \citep[see][for the case of $\omega$\,CMa]{ghoreyshi2021}. 
To calculate the expected observables of our SPH models, we used HDUST, {following the procedure detailed in \citet{suffak2024}}. By qualitatively comparing the models with the data presented by Paper I, we seek to constrain the mass injection parameters to physically plausible and practically useful ranges.

\paragraph{Photometry}

The evolution of the simulated signals in photometry (in the TESS band -- from 600 -- 1000 nm), H$\alpha$ EW, \ew, $\mathrm{PS_W}$ and $B$-band polarisation are presented in Fig.~\ref{fig:besthdust}. 
The different colours represent the inclination angle of the rotational axis of the Be star with respect to the observer: at 0$\degree$ the star is seen pole-on, while at 90$\degree$ it is edge-on. 
{Our template, f\,Car, has an intermediate inclination angle of about $66\degree\pm16\degree$ \citep{zorec2016}. We calculated HDUST simulations for our SPH models with $i = 50\degree$, $85\degree$ and $20\degree$ to analyse the effect of inclination on the observables, and to allow for comparison with observations of other Be stars.}

\begin{figure}
    \centering
    \resizebox{\hsize}{!}{\includegraphics{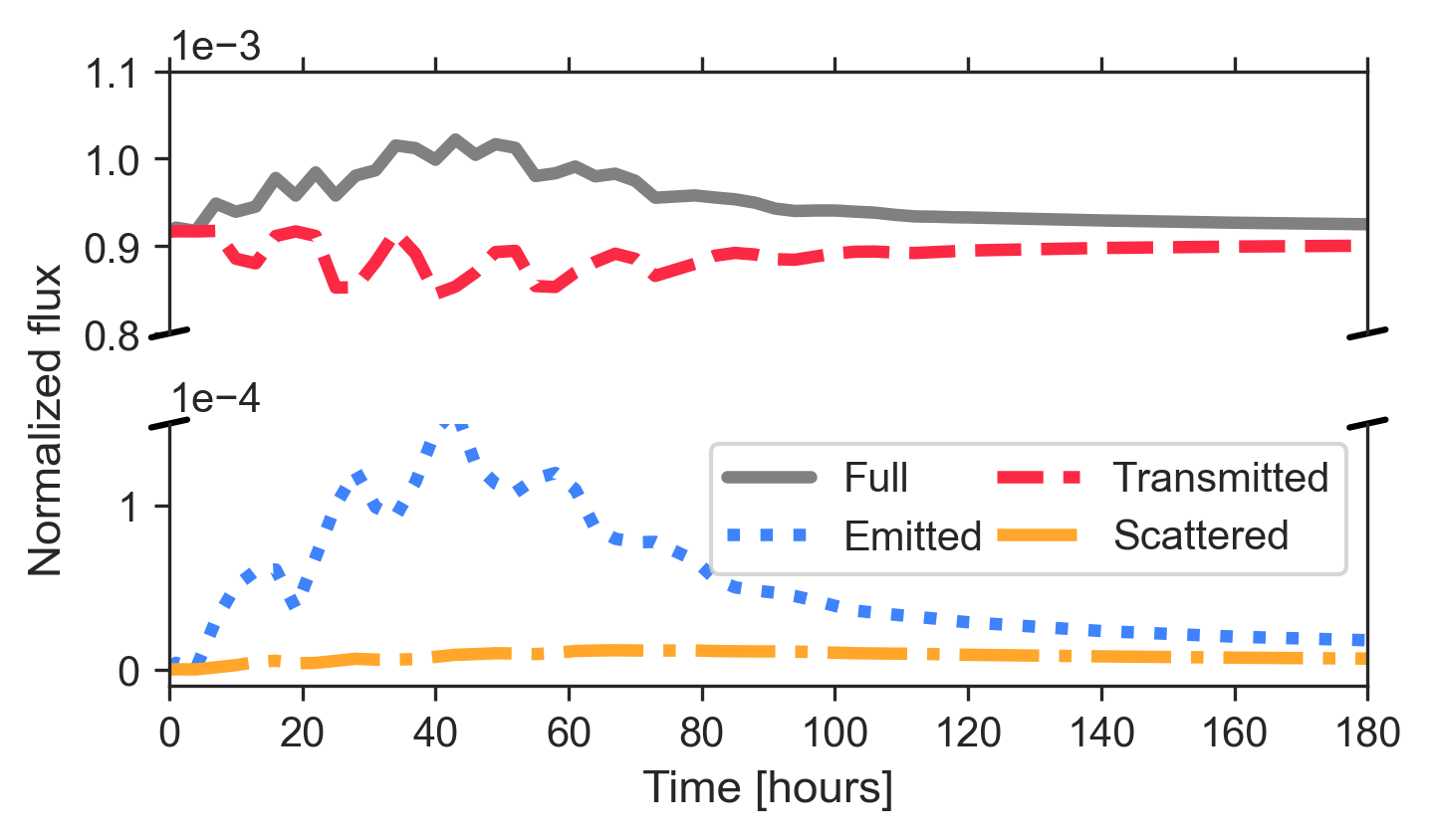}}
    \caption{{Emitted, transmitted, scattered and full flux as calculated by HDUST for our preferred model, seen at 85$\degree$. The photometric oscillations present in the transmitted flux (i.e. flux that comes directly from the star, unimpeded) dictate the oscillations in the full flux. We note the jump in scale on the y-axis, as the full and transmitted fluxes are one order of magnitude larger than the emitted and scattered fluxes.}}
    \label{fig:flux_ess}
\end{figure}

The shape of the photometric curves of our simulation (top left panel of Fig.~\ref{fig:besthdust}) is qualitatively very similar to the `dips' and `bumps' seen in observational data of outbursting Be stars. Whether there is an increase or a decrease in continuum emission {following an outburst}
depends on the inclination angle: pole-on Be stars will show bumps, as the net emission of the system increases when hot material is injected into orbit, while edge-on Be stars have dips, as the dense material obscures part of the star \citep[see][for detailed descriptions of these phenomena]{harmanec1983, haubois2012}. {The light curves of the model seen at 85$\degree$ (and faintly at 50$\degree$) inclination angle have periodic photometric decreases consistent with {transits or partial `eclipses'} caused by the asymmetric structure orbiting the star. As it circularises, the effect diminishes, disappearing around 100\,h}. To investigate the source of this observational effect, we analyse the relative values for the emitted, scattered and transmitted fluxes of the star+disk system, as calculated by HDUST (Fig.~\ref{fig:flux_ess}). The photometric oscillation seen in the integrated light is identical to the signal observed in the transmitted flux, which measures the fraction of starlight that reaches the observer unimpeded. Given that the transmitted flux level is much higher than the emitted level {(which measures the light emitted by the disk)} ---as expected by the low photometric amplitude of the flicker---, this clearly indicates that the oscillations seen at higher inclinations are caused by density enhancements in the disk transiting in front of the star. {Further corroboration is that the emitted and transmitted flux variations are half an orbital cycle out of phase. This happens because, while the transiting density enhancement blocks starlight, attenuating the transmitted flux, it also re-emits light, thus the enhancements in the emitted flux (at a much lower amplitude).} {This effect has not been directly observed in outbursting Be stars; it might be due to it being disguised by other oscillatory signals (due to stellar pulsation, for instance) that are not included in our simulation, or it is an indication that real Be outbursts are even more concentrated in the equator than we assume in our simulations. }

The general behaviour of the simulated photometry, apart from the partial eclipses, is similar in all inclinations. There is a fast and constant increase (decrease, for 85$\degree$) during outburst, {and the peak of the signal is reached a few hours after mass ejection stops: between 45-55\,h in all inclinations.}
From this point onward, the signal decreases with a slope that becomes shallower with time, returning to the baseline (within 0.5\% variation) at about {180\,h for 85$\degree$, 200\,h for 50$\degree$, and 300\,h for 20$\degree$} {(at lower inclination angles, the emitting area of the disk is larger, thus the emission endures for longer)}. This result is consistent with the observed behaviour of flickers in that dissipation takes longer than build-up. In the observations, the dissipation time of their photometric signals is roughly twice as long as the build-up time \citep{labadie-bartz2018,figueiredo2025}. For our example, f\,Car, which has an inclination angle between $\sim 50-70\degree$, the duration of the build-up is about 80\,h, while the dissipation is estimated to end at around 200\,h, 120\,h later (see Fig.~\ref{fig:fcar}), considering a 0.5\% variation from the baseline, defined from the non-outbursting data in the light curve.

We measured the slopes of the build-up and dissipation parts of the curve for the three inclinations at 20\,h and 70\,h. {Results are shown in Fig.~\ref{fig:mags_slopes} and summarised in Table~\ref{tab:mags_slopes}}. {Our simulation at 50\degree has slightly steeper build-up and dissipation slopes than f\,Car (see also Fig.~\ref{fig:mags_slopes}) while also reaching a similar peak signal of about -0.1 mag. These values are overall consistent with the flickers presented by Paper I.}

\begin{table}[]
    \centering
    \caption{Slopes of build-up and dissipation of the flickers in our preferred model and in f\,Car.}
    \begin{tabular}{c|c|c}
    \hline
    Inclination & Build-up slope & Dissipation slope \\
     & [$\rm mmag \, h^{-1}$] & [$\rm mmag \, h^{-1}$] \\
    \hline
    20\degree & 4.0 & -4.9\\
    50\degree & 2.2 & -2.6\\
    85\degree & 0.9 & -0.5\\
    \hline
    f\,Car ($\sim$ 50\degree) & 1.8 & -2.0\\
    \hline
    \end{tabular}
    \tablefoot{Build-up is measured at 20\,h and dissipation at 70\,h. See also Fig.~\ref{fig:mags_slopes}. }
    \label{tab:mags_slopes}
\end{table}

\paragraph{H$\alpha$ emission}

{The H$\alpha$ line profiles of our model are visually very similar to the line profiles of f\,Car (Fig.~\ref{fig:fcar}) in shape and emission intensity (the four right-hand panels of Fig.~\ref{fig:besthdust}). We further scrutinise the behaviour of the H$\alpha$ emission by measuring the variation in equivalent width ($\Delta \rm EW$), oscillations in \ew and $PS_W$ (left-hand panels of Fig.~\ref{fig:besthdust}).}

In f\,Car, the EW {reaches a minimum}\footnote{By definition, EW decreases when the relative emission-line strength increases.} at $\sim$110\,h, about 30\,h after the photometry, and the decrease in emission proceeds very slowly (Fig.~\ref{fig:fcar}). For the simulations, the dip in EW (second panel of Fig.~\ref{fig:besthdust}) is not as clear as f\,Car's, but a minimum is reached at around 100\,h, 50\,h after the peak in photometry, followed by an extended plateau. {The largest variation in EW value in the models is around -2\,\AA, the same as in f\,Car.}

Density asymmetries in the disk cause variations in the \ew ratio of the H$\alpha$ line (third panel). As such, these variations dampen out as the material circularises. Indeed, the amplitude of the variations decreases greatly by $\sim$80\,h, but persists for many days after the end of mass injection. This enduring signal might originate from the outermost region of the disk, where the density still has significant azimuthal dependence (see the map in the bottom panel of Fig.~\ref{fig:bestsph}); however, given the stability of the frequency, the reaccretion of the material from the innermost disk also play a role. {The largest amplitude of the \ew oscillations is seen at 50\degree, due to the combination of a large projected emitting area and partial obscuration of the disk by the central star}. 

There is a noticeable phase shift in the oscillations happening during the transition in mass ejection state at 42.4\,h. Whether a similar shift is present in the observational data of Paper I is not easily verifiable, as the data is too sparse, but it was observed in V537 Lac (see Appendix B.5 in Paper I), attributed to two flickers happening in the span of a few days (a `compound outburst'). 
We note that our simulations impose an brusque cut-off to the mass ejection, which may be responsible for the observed discontinuity. In nature, it is unlikely that the mass ejection ends as abruptly. 
{In spite of the phase shift, the oscillatory frequency changes little between the two states for the model seen at 50\degree. Following the same procedure of Paper I (their Sect.\,3.6), these frequencies are 1.48 and 1.50 $\rm c\,d^{-1}$ respectively. At 20\degree, similar frequencies are found, while at 85\degree, no good fit to a sinusoidal function could be derived.}

The $\rm PS_W$ (Eq.~\ref{eq:psw}) also offers insight into the evolution of the distribution of material around the star (fourth panel of Figs.~\ref{fig:besthdust} and \ref{fig:besthdust_zoom}). The value peaks only at around 100\,h, about two days after the end of mass injection. {Compared to f\,Car, $\rm PS_W$ in our simulation returns to its pre-outburst level much faster: the slope of the 50$\degree$ curve at 200\,h is $72 \, \rm km\,s^{-1} h^{-1}$, while the data decreases with a slope three times smaller, $25 \, \rm km\,s^{-1} h^{-1}$.}

In the observational data for f\,Car, the curves for EW and $\rm PS_W$ mirror each other with very similar timescales and shapes. These two observables behave differently in the simulations, however. 
The EW is a measurement connected to the size of the emitting area, thus it is very sensitive to the volume, mass and opacity of the orbiting material, and the rate of decay of the emission depends strongly on the viscosity. {The $\rm PS_W$, on the other hand, is more sensitive to the shape of the line profile itself, thus likely linked to the detailed kinematics of the circumstellar material.}
In our simulation, the decay of H$\alpha$ EW is much slower than in the data. Previous works comparing dissipating Be decretion disk models to spectra have encountered similar difficulties. 
\citet{ghoreyshi2021} finds that increasing the viscosity provides a better agreement between the EW of the Be star $\omega$\,CMa to their models of dissipating disks; however, increasing the viscosity throughout the whole disk causes the simulated photometric light curve to decrease too fast. Since H$\alpha$ probes a larger emitting area than the visible bands, the authors suggest that a radially variable viscosity could balance the two timescales. By having the viscosity increase with radius, the dissipation of the outer disk (or outer proto-disk, in our case) and, consequently, the decrease in H$\alpha$ emission would be faster, while the inner disk (the region probed by the visible bands and where the models and data are already in good agreement) would remain unchanged. Another phenomenon that could cause faster dissipation of the disk is pressure from stellar radiation \citep{kee2016}, which is not considered in our simulations.

\paragraph{Polarimetry}

Finally, we also investigate the $B$-band polarimetric signal (fifth panel of Figs.~\ref{fig:besthdust} and ~\ref{fig:besthdust_zoom}). For the low inclination of 20$\degree$, as expected, the signal only reaches about 0.05$\%$ at its maximum, increasing to nearly 0.3$\%$ for the more edge-on case of 85$\degree$. The polarimetric peak is reached at 62\,h for 50$\degree$ inclination (20\,h after the end of the outburst and 40\,h before the lowest point in EW) and at 70\,h for 85$\degree$ (28\,h after the end of the outburst and 30\,h before the EW dip). Similarly to the photometric curve, the signal decreases with a slope that becomes shallower over time. And similarly to the H$\alpha$ emission, the signal takes many days to disappear, returning to a baseline on a timescale slower than photometry, but faster than H$\alpha$. Unfortunately, a direct comparison to f\,Car is not possible, as no polarimetric monitoring was conducted during its outburst phase. 
{However, recent data on another star studied in Paper I, 12\,Vul (which has similar H$\alpha$ EW variations amplitudes and inclination angle as f\,Car),} report polarimetric variability of about 0.2\%, in very good agreement with our model predictions \citep{carciofi2025}. Similar polarimetric amplitudes were reported for a mass outburst of the Be star $\alpha$\,Eri \citep{carciofi2007}.

In conclusion, considering the simple assumptions for the mass ejection made in our SPH simulations, the agreement with the data is encouraging. We recover similar shapes and amplitudes in photometry, and similar frequencies in the \ew variations compared to the observations of f\,Car, {one of} the most pristine and well-documented example of an outbursting Be star to date, considering cases with full coverage in both space photometry and time-series spectroscopy. 
This agreement highlights the ability of the VDD framework to capture the essential physics of Be disk formation, provided a more realistic mass injection is assumed.
In the subsections that follow, we investigate the effects of the key parameters in our simulations and justify our choice for the model detailed above (referred to as our `preferred model' from this point onwards) as the closest representation for the template f\,Car flicker.

\subsection{Sub- vs. super-Keplerian mass injection}\label{sec:kepl}

\begin{figure}[h]
    \centering
    \resizebox{\hsize}{!}{\includegraphics{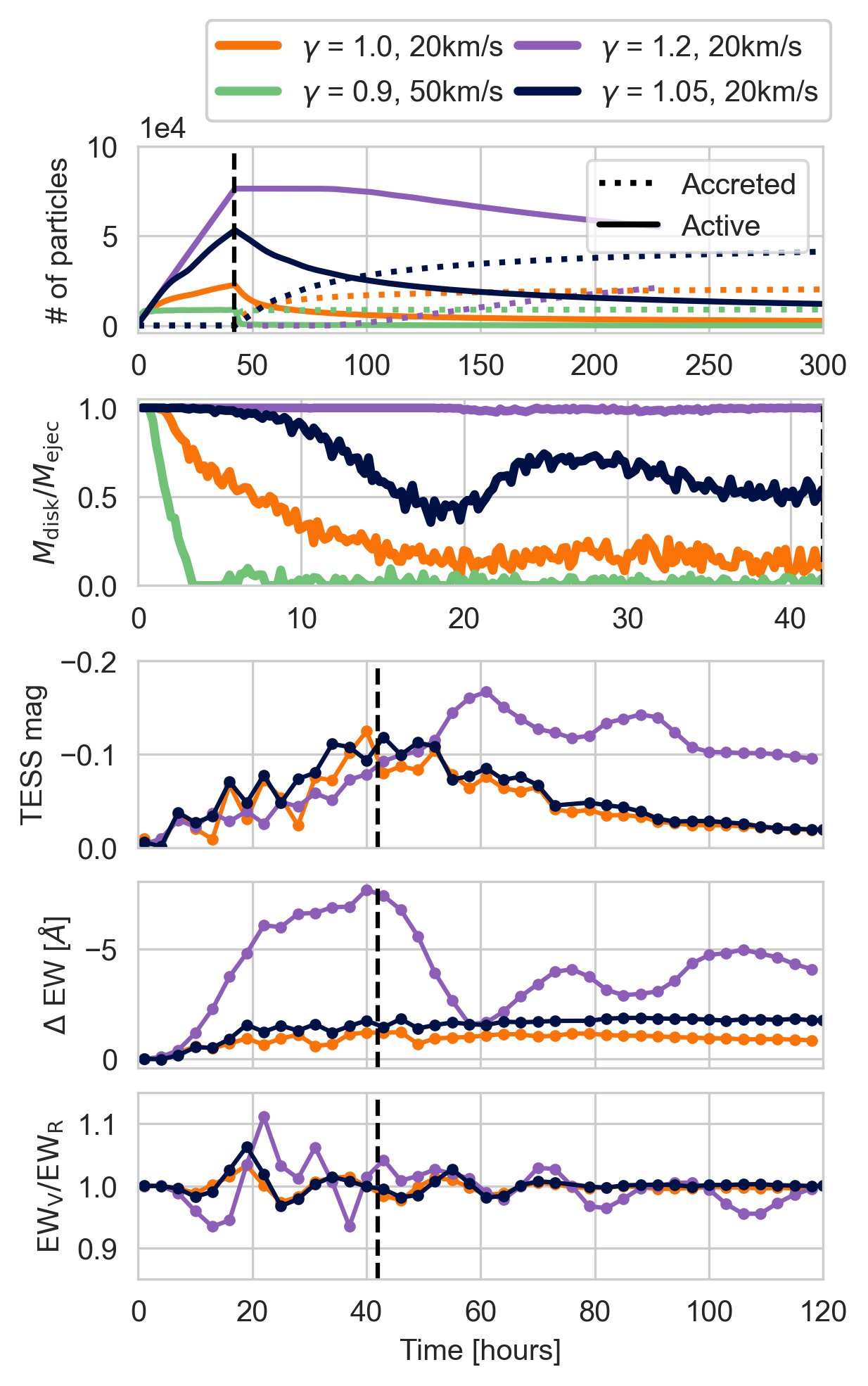}}
    \caption[Comparison between models with sub-Keplerian ($\gamma < 1.0$), Keplerian ($\gamma=1$), and super-Keplerian ($\gamma>1.0$) mass injection.]{Comparison between models with sub-Keplerian ($\gamma < 1.0$), Keplerian ($\gamma=1$), and super-Keplerian ($\gamma>1.0$) mass injection. The top panel compares the number of particles injected and accreted. The second panel shows the efficiency of mass decretion: {a value of 1 indicates that all ejected particles remain in orbit; conversely, a value of zero means complete reaccretion}. The three bottom panels compare HDUST synthetic observables for the $\gamma=1$ and $\gamma>1$ cases: photometry (third panel), EW (fourth panel) and \ew (fifth panel), for an inclination angle of 50$\degree$. The dashed black line marks 42.4\,h, the end of mass injection in all models.}
    \label{fig:comp_small}
\end{figure}

The dynamics of the ejecta is, as expected, highly dependent on the rotational velocity of the injection volume (\gam) and on the viscosity ($\alpha$), while the isotropic velocity ($v_i$) and the dimensions of the injection volume play a comparatively minor role. In this section, we focus on the effects of \gam and $v_i$ in our SPH simulations and explore the influence of the remaining parameters in the sections that follow.

We ran simulations with injection volumes rotating at sub-Keplerian, Keplerian, and super-Keplerian velocities. All our SPH simulations were calculated with a fixed {ejection} rate, $\dot{M}_{\rm ejec} = 10^{-7} \, \rm \, M_\odot \, yr^{-1}$. This value is commonly used in SPH simulations of Be stars since \citet{okazaki2002}, as it was the rate necessary to produce sufficiently massive disks. As all particles have the same mass and the material is not self-gravitating, this mass ejection rate can be arbitrarily rescaled when computing observables with HDUST. Accordingly, the density of each model was adjusted up or down to approximately match the observed amplitudes in brightness and H$\alpha$ EW seen in the data of f\,Car. In practice, this means that each HDUST simulation has its own `effective' mass ejection rate. For our preferred model, the observables shown in Figs.~\ref{fig:besthdust} and \ref{fig:comp_small} use a scaled mass ejection rate of $\dot{M}_{\rm ejec}^{\rm scaled} = 1.0 \times 10^{-6} \, \rm M_\odot \, yr^{-1}\, \mathrm{str}^{-1}$ (that is, the density is lowered by 40\% from the original SPH results), considering the solid angle of the injection volume $\theta = 0.04 \rm \, str$.

Not all of the ejected material is truly integrated into the disk. The ratio between the amount of material that enters into an orbit and the material the Be star is ejecting is a measure of the efficiency of mass decretion: a value of 1.0 indicates that all material the star ejects reaches an orbit, {while a value of 0.0 indicates that all ejected matter has fallen back to the star (second panel of Fig.~\ref{fig:comp_small} and top panels of Figs.~\ref{fig:comp_big} and \ref{fig:comp_big}).} Therefore, the {disk growth rate}, $\dot{M}_{\rm disk}$, is always a fraction of $\dot{M}_{\rm ejec}^{\rm scaled}$. As an example, our preferred model has a mean efficiency of 69$\%$, thus $\dot{M}_{\rm disk} = 6.9 \times 10^{-7} \, \rm M_\odot \, yr^{-1} \, \mathrm{str}^{-1}$.

The sub-Keplerian ($\gamma = 0.9$, in green) simulation fails to lift much of the ejected material from the injection volume into orbit, even with a $|v_i| = 50 \, \rm km\,s^{-1}$, well above the sound speed expected in the stellar photosphere ($c_s \approx 20\, \rm km\,s^{-1}$). Thus, the outburst in these conditions produces only minor effects: the material injected has insufficient AM to sustain orbit or expand, and falls back to the star almost immediately. The ineffectiveness of the decretion in this simulation is clear, as the mean decretion efficiency drops rapidly only about 2\,h after the start of mass injection, {and} remains close to zero until it terminates at 42.4\,h. Thus, calculating the HDUST observables would require a larger scaling to increase the density, placing $\dot{M}_{\rm disk}$ in the order of $10^{-4} \, \rm M_\odot \, yr^{-1}\, \mathrm{str}^{-1}$, or larger. Even then, this model would no be able to reproduce the behaviour of the data because the emitting area would still be too small.

The simulation with $\gamma = 1.0$ (in orange), for which the angular frequency of the injection volume is exactly the orbital angular frequency at the stellar surface, and $|v_i| = 20 \, \rm km\,s^{-1}$ is able to inject more material into orbit than the sub-Keplerian cases, but is still not very effective. For this model, in average, only 28\% of the material ejected remains in the disk at the end of the build-up phase. 
To achieve a similar photometric variation to f\,Car, the $\gamma=1.0$ model requires a mass ejection rate 10 times larger than the $\gamma = 1.05$ model, i.e. $\dot{M}_{\rm ejec}^{\rm scaled} = 1 \times 10^{-5}\, \rm M_\odot \, yr^{-1}\, \mathrm{str}^{-1}$ and an $\dot{M}_{\rm disk} = 2.9 \times 10^{-6}\, \rm M_\odot \, yr^{-1}\, \mathrm{str}^{-1}$. The emission strength, however, only reaches -1\AA.

\begin{table*}[]
    \centering
    \caption{Mean decretion efficiency, scaling factor for the HDUST calculations, scaled mass decretion and AM rates for the models with $\gamma = 1.0$, 1.05 and 1.2 (Fig.~\ref{fig:taucompare}).} 
    \begin{tabular}{ccc|cc}
    \hline
     Model & Mean decretion eff. & Scaling factor & $\dot{M}_{\rm disk}$ & AM dec. rate \\
     $\gamma$ & [$\dot{M}_{\rm disk}/\dot{M}_{\rm ejec}$] & (for HDUST models) & [$\rm M_\odot \, yr^{-1}\, \mathrm{str}^{-1}$] & [$\rm g \, cm^2 \, s^{-2} \, str^{-1}$]\\
    \hline
     1.0   & 28\% & 4.0 & $2.9 \times 10^{-6}$ & $3.8 \times 10^{39}$\\
     1.05  & 69\% & 0.4 & $6.9 \times 10^{-7}$ & $9.2 \times 10^{38}$ \\
     1.2   & 98\% & 0.2 & $4.9 \times 10^{-7}$ & $6.5 \times 10^{38}$\\
    \hline
    \end{tabular}
    \label{tab:rates}
\end{table*}

We also calculated a model where the rotation of the injection volume is significantly larger than Keplerian, with $\gamma=1.2$. This model forms a much larger disk than its predecessors, as it places material into orbit more effectively (purple line in the second panel of Fig.~\ref{fig:comp_small}). In fact, mass decretion is so efficient that the density had to be scaled down for the HDUST calculations, with an $\dot{M}_{\rm ejec}^{\rm scaled} = 5 \times 10^{-7} \, \rm M_\odot y^{-1} \mathrm{str}^{-1}$.
The shapes of photometric and EW curves for this simulation are clearly very different from those observed in f\,Car and other outbursting Be stars. The complex variations in EW are particularly striking. Because the injected material carries a large amount of specific AM, it rapidly reaches large orbits, causing the disk to initially expand quickly in size, but not in density. This explains the rapid increase in optically thick H$\alpha$ line emission, contrasted with the slower rise in optically thin TESS continuum emission (see purple lines in Fig.~\ref{fig:taucompare}). The continuum emission becomes significant only after the end of the mass injection phase, when part of the material, having lost angular momentum through shear interactions, falls back onto the star. This fallback leads to a rise in emission, as seen around 60\,h in the third panel of Fig.~\ref{fig:comp_small}. Such complex behaviour is not observed in the data, suggesting that extremely high rotation rates in the injection volume are unrealistic. These differences also cannot be compensated by changing $\dot{M}_{\rm ejec}^{\rm scaled}$, as the $V$-band and H$\alpha$ emitting areas would require different densities to match the data.

The values for the scaling factor and mass decretion rates of the three of the models ($\gamma$ = 1.0, 1.5 and 1.2) are summarised in Table~\ref{tab:rates}. We also estimate the AM decretion rate, given $\dot{M}_{\rm disk}$ and the angular velocity of the particles. \citet{rimulo2018} estimated AM-loss rates around $\sim 5 \times 10^{36} \, \rm g \, cm^2\, s^{-2}$, based on the disk events of Be stars in their sample. Again assuming AM is lost from an area of similar size {to that of} our simulations, the value becomes $\sim 1.25 \times 10^{38} \, \rm g \, cm^2\, s^{-2} \, str^{-1}$. The implications of the values for the mass and AM decretion rates are discussed in Sect.~\ref{sec:discussion}.

In conclusion, our results show that 
$1.0 \leq \gamma < 1.2$ is necessary for the formation of a disk structure that 1) has a large enough emitting area in both visible and H$\alpha$ wavelengths, and 2) evolves on a similar timescale as the observational data. 
We find that the \gam = 1.05 model, although broadly similar to the denser \gam = 1.0 model, provides a better overall match to the observations once all diagnostic details are considered.

\subsection{Effects of the opening angle}\label{sec:op_ang}

\begin{figure*}
    \centering
    \resizebox{\hsize}{!}{\includegraphics{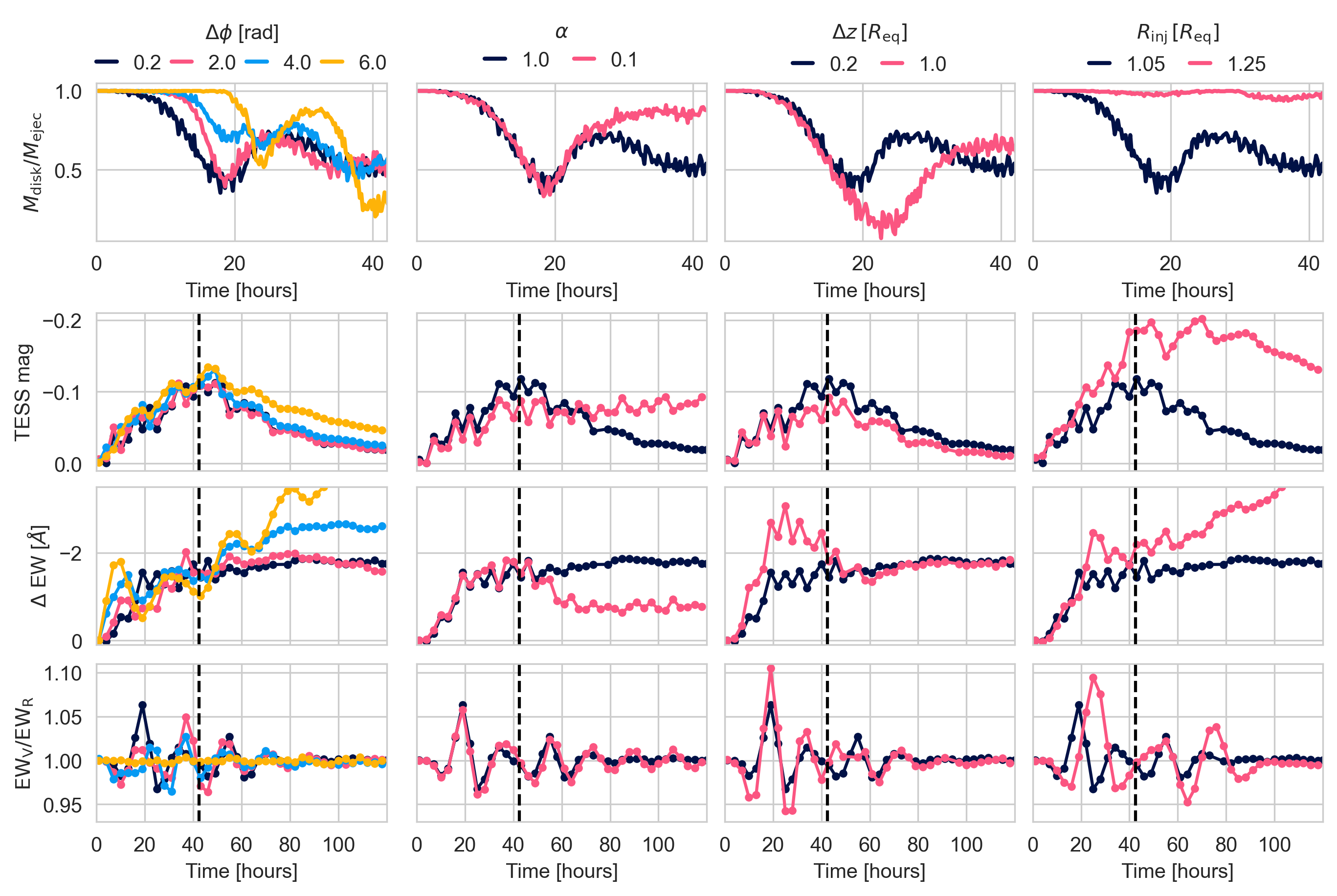}}
    \caption[Comparison between models with  $\gamma=1.05$, but varying $\Delta\phi$, $\alpha$, $\Delta z$, and $R_{\rm inj}$.]{Comparison between models with $\gamma=1.05$, but varying the azimuthal extent of the injection volume ($\Delta\phi$) in the first column, the viscosity parameter $\alpha$ in the second column, the vertical height $\Delta z$ in the third column, and the injection radius $R_{\rm inj}$ in the fourth column. The model in dark blue in all panels is our preferred model, described in detail in Sect.~\ref{sec:most_fave}. The top panel shows the efficiency of mass decretion {(calculated from 0 to 42.4 h, when mass ejection is active)}. The four bottom panels compare HDUST synthetic observables {during the first 120\,h of the simulation}: photometry (second panel), EW (third panel) and \ew (fourth panel) for an inclination angle of 50$\degree$. The dashed black line marks 42.4\,h, the end of mass injection in all models.}
    \label{fig:comp_big}
\end{figure*}

The strongest evidence that Be stars lose mass from a confined region rather than along their entire equatorial circumference comes from the cyclic V/R variations in H$\alpha$ and other emission lines observed concurrently with photometric flickers. 
We explored four values of the opening angle, $\Delta \phi$, for the injection volume to determine whether the observables, in particular the \ew amplitude, could be used to derive an upper limit. The first column of Fig.~\ref{fig:comp_big} compares these four simulations, where all parameters, except $\Delta \phi$, are fixed to the values of our preferred model ($\Delta \phi = 0.2$, in dark blue): $\Delta \phi = 2.0$ in pink, $\Delta \phi = 4.0$ in light blue, and $\Delta \phi = 6.0$ in yellow. The inclination angle of the observer is 50$\degree$. Results for an edge-on orientation are shown in Fig.~\ref{fig:comp_big_85}.

The shape of the photometric flickers (second row of Fig.~\ref{fig:comp_big}) changes little when $\Delta \phi$ is increased from 0.2 to 2 radians (dark blue and pink lines). This holds for all inclination angles. The EW also behaves similarly but shows an increase in emission. The frequencies and amplitudes of \ew, and the behaviour of $\rm PS_W$ are also very similar to the preferred model.

Increasing the opening angle of the injection volume further, we have models with $\Delta \phi$ = 4.0 and 6.0 rad (i.e. an injection volume that covers most of the equator of the star), in light blue and yellow. These two models already show some characteristics not present in either the data or the models with smaller $\Delta \phi$. 
{At near edge-on inclination (Fig.~\ref{fig:comp_big_85}), we no longer see the eclipses caused by the concentrated material orbiting the star when $\Delta \phi = 6.0 \, \rm rad$ (as discussed in Sect.~\ref{sec:most_fave}), as the injection volume is now much wider in azimuth. Conversely, bumps are still present, and more prominent, in the light curve at 50$\degree$. This behaviour is caused by collisions between material expelled in a previous orbital cycle and freshly injected material, an effect that grows in intensity with increasing opening angle. This expansion/contraction of the disk is not visible when the system is viewed edge-on, thus the absence of this effect in the light curve for $\Delta \phi = 6.0$ rad in Fig.~\ref{fig:comp_big_85}.}
The \ew variations have much smaller amplitudes for the models with large $\Delta \phi$. The frequency of the variations for model $\Delta \phi$ = 4.0 rad is not dissimilar to those of models with smaller $\Delta \phi$, but model $\Delta \phi$ = 6.0 rad has a more erratic behaviour. 
{The $\rm PS_W$ mirrors the EW quite closely, including the bumps and the more pronounced decrease with increasing $\Delta\phi$.}

Considering all these observational characteristics, we conclude that Be stars must lose mass from a region with an azimuthal extent $\lesssim 4 \, \rm rad$, i.e. covering less than $\sim60\%$ of the stellar equator. It should be pointed out, however, that the models only offer a loose constraint on this parameter.

\subsection{Effects of viscosity}

During the outburst, the role played by viscosity is nearly negligible, as the dynamics of the material are dominated by the influx of AM rich material from the injection volume. After the end of mass injection, the spreading of AM and the subsequent rate of growth of the disk are dictated by viscosity. The second column of Fig.~\ref{fig:comp_big} (and \ref{fig:comp_big_85}) compares our preferred model, with $\alpha = 1.0$ (dark blue line), to a model with the same parameters, but with $\alpha = 0.1$ (pink line). In photometry (second panel), the behaviour of the flicker is nearly identical between the two models until mass injection is terminated at 42.4\,h. From this point onward, the observables of the low $\alpha$ model begin to diverge. 
As this model is less effective in spreading the material, it remains concentrated in lower orbits for a much longer time than in the high $\alpha$ model. Consequently, the light curve has a different shape, plateauing after mass injection terminates, persisting past 120\,h.
For the near edge-on case (Fig.~\ref{fig:comp_big_85}), the slower circularisation and radial spread also leads to a longer duration of the variations in photometry caused by the partially eclipsing dense material, and to {less H$\alpha$ emission}. 
Additionally, the damping of the \ew variations takes longer than in the high-$\alpha$ model, although the frequency remains constant.

{The timescale of the dissipation of the disk in the low-$\alpha$ model is much longer than observed in the data, producing incompatibility especially in the light curve evolution. 
Thus, we are motivated to prefer higher viscosity values.}
We note, however, that out treatment of viscosity in the SPH simulations has relevant approximations, as discussed in Sect.~\ref{sec:results}.

\subsection{Effects of vertical height}\label{sec:height}

We explored two values for $\Delta z$, the vertical height of the injection volume from the equatorial plane: 0.1 (in our preferred model) and 0.5 \req.
While mass injection is active, there are particles at all heights because their initial positions in all coordinates are randomly assigned. As the simulation evolves, they quickly move to lower $z$, around the equator. In an $\alpha$-disk as set up in our SPH simulations, where the disk is isothermal and in hydrostatic equilibrium, its scale height $H$ is controlled by the gravity of the star and the disk gas pressure,

\begin{equation}
    H = H_0 \left(\frac{r}{R_{\rm eq}}\right)^{3/2}
\end{equation}

where $H_0 = c_s \, R_{\rm eq}/v_{\rm orb}$. Combined with the effect of the conservation of angular momentum, this initial distribution starts to collapse as soon as mass injection stops, oscillating around $z = 0$ (see the evolution of the scale height in the lower panels of Fig.~\ref{fig:eccs}). The subsequent disk evolution is similar to that of our preferred model, with $\Delta z = 0.2$\req (see Fig.~\ref{fig:tall}), but the disk takes a longer time to circularise.

Observationally, the largest effect of this increased vertical height is in H$\alpha$. The larger emitting area during mass injection translates into larger {emission} and larger shifts in \ew during the mass ejection phase of the simulation than in our preferred model (third column in Figs.~\ref{fig:comp_big} and ~\ref{fig:comp_big_85}, for 50 and 85\degree inclinations). This hump+plateau pattern in EW is not seen in the observations presented in Paper I, and thus we conclude that mass ejection happens over a limited range in latitude around the stellar equator.

\subsection{Surface vs. lever-arm injection}\label{sec:lever}

We compare our preferred model with one for which the centre of the injection volume is 0.25~\req above the stellar equator in the fourth column of Fig.~\ref{fig:comp_big}. This model investigates the idea proposed by \citet{nixon2020}, where the material is given magnetic torque via the inner disk boundary. The matter and AM are added to the disk a radius slightly larger than the corotating radius. {The rotation of the injection volume and the particles have, as in our preferred model, $\gamma = 1.01$. Considering the higher injection radius, the corresponding velocity at the equator is $\Omega/\Omega_{\rm crit} = 0.75$}. The immediate effect is that the decretion efficiency (first panel) is nearly 100\%: virtually all of the ejected material reaches orbit successfully. As these two models are scaled to the same mass decretion rate, this translates to a larger emitting area in both photometry and H$\alpha$ (second and third panels). The \ew variations have a lower frequency than in our preferred model (fourth panel).

\begin{figure}
    \centering    
    \resizebox{\hsize}{!}{\includegraphics{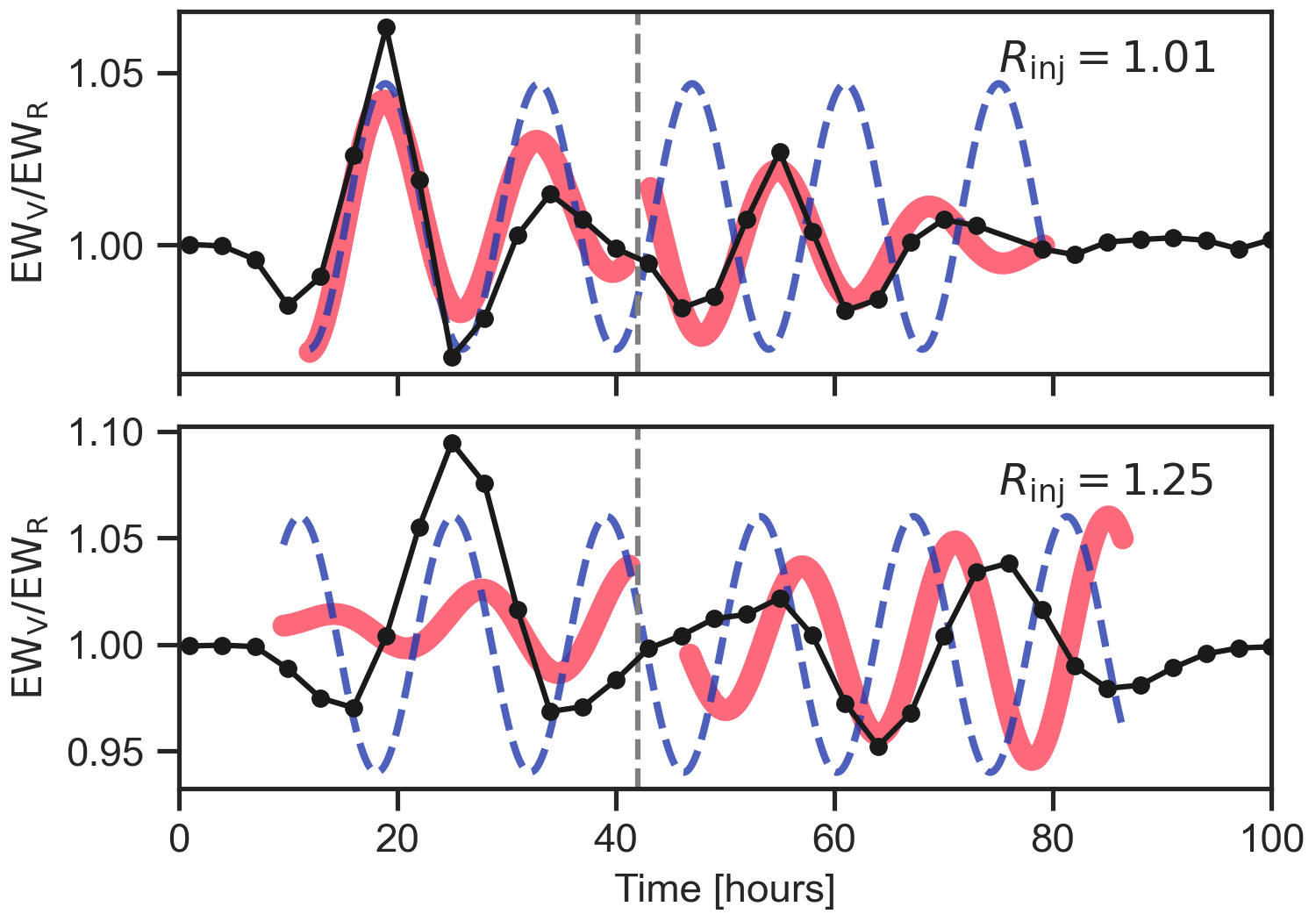}}
    \caption[\ew variations for the model with $\gamma=1.05$]{\ew variations for the model with $\gamma=1.05$, $\alpha=1.0$ and $R_{\rm inj} = 1.01$ (top panel), and a model with the same parameters, but with $R_{\rm inj} = 1.25$ (bottom panel), for 50$\degree$ inclination. In both panels, the coloured lines represent comparisons with a sinusoidal function (Eq.2 of Paper I) with a frequency of 1.71 $\rm c \, d^{-1}$, the orbital frequency of f\,Car. The red lines show the best solutions we found when analysing the build-up phase (from 5 to 42\,h) and the dissipation phase (from 43 to 90\,h) separately. The dashed blue line shows the sinusoidal function for both phases combined (from 5 to 90\,h), ignoring the phase jump that happens at the end of mass injection. }
    \label{fig:freq}
\end{figure}

Paper I finds a nearly 1:1 correlation between the \ew frequencies and the orbital frequency at the stellar equator 
\begin{equation}
    f_{\rm orb} = \frac{1}{2\pi} \sqrt{\frac{GM}{R_{\rm eq}^3}}\,,
\end{equation}
for all stars in the sample, strongly suggesting that the emitting material is orbiting either at or very near the stellar equator. This value was estimated to be $1.71 \pm 0.47 \rm \,c\,d^{-1}$ for f\,Car, assuming $M = 7.60 \pm 0.30 \, \rm M_{\odot}$ and $R_{\rm eq} = 5.8 \pm 1.3 \, \rm R_{\odot}$ \citep{zorec2016}, matching the $1.72 \rm \,c\,d^{-1}$ \ew frequency.

{
As we used the same stellar parameters in our simulations, we also expected to recover this near 1:1 correlation. 
For our preferred model, a good agreement between the model and the sinusoidal curves with $1.71 \rm \,c\,d^{-1}$ frequency (Eq. 2 from Paper I) can be reached if we consider the injection/build-up phase (from about 5 to 42 h) separately from the remainder of the simulation (see the two separate red curves), as there is a phase shift between the two (Fig.~\ref{fig:freq}, top panel). In contrast, the lever-arm model (bottom panel), having a visibly lower frequency, cannot reproduce the sinusoidal curve in either build-up or dissipation. This follows from the fact that the material is in a higher orbit. }

While the higher photometric signal and H$\alpha$ excess (second and third panels of the fourth column in Fig.~\ref{fig:comp_big}) could be comparable with f\,Car if the $\dot{M}_{\rm inj}^{\rm scaled}$ was decreased, the frequency of the \ew would remain the same. However, the 1:1 correlation between $f_{\rm orb}$ and the \ew oscillation frequency found in Paper I rests on the stellar parameters assumed for the central star. The equatorial radius, in particular, has an uncertainty of $1.3\, R_{\odot}$, a much larger margin than our range of \rinj. Considering this, the present analysis cannot wholly exclude this mass injection scenario.

\section{Discussion and conclusions}\label{sec:discussion}

We explored a simple mass ejection scenario for Be stars, where material is lost from an injection volume that rotates around the stars with an added ballistic velocity of random orientation {and fixed modulus} in the stellar corotating frame. We used the pristine outburst of f\,Car captured by TESS and NRES in Paper I, between dates 3010 and 3030 TESS JD, as the basis for our comparison.

Our results show that the amplitudes and timescales of the photometric and H$\alpha$ variations of the reported outbursts are qualitatively compatible with a mass ejection event where:

\begin{enumerate}
    \item Ejecta velocity: models where mass is injected with rotational velocity at or below the Keplerian value (i.e. equal or smaller than $(GM/R_{\rm inj}^3)^{1/2}$) only create disks with sufficient mass to reproduce the observational signals of Be outbursts if their mass ejection rate is scaled to values similar to the wind mass-loss rates of the much hotter O-type stars (Sect.~\ref{sec:kepl}). Thus, we propose that material is injected with a slightly super-Keplerian velocity; 
    \item Azimuthal opening angle of the injection zone: to reproduce the characteristic oscillations in the two peaked profiles of H$\alpha$ emission lines seen in the early stages of outbursting Be stars, material cannot be lost from the entire equatorial circle of the star; in Sect.~\ref{sec:op_ang}, we show it cannot be larger than $\sim4$\,rad, or $\sim 64\%$ of the equator;
    \item Viscosity: the observables of models with high viscosity ($\alpha = 1.0$) evolve on timescales comparable to what is observed in the data, in contrast with the much slower timescales of low $\alpha$ models;
    \item Vertical height of the injection zone: while this parameter was not explored in great depth, increasing it leads to a `double hump' pattern in the equivalent width of H$\alpha$ that is not present in any observations. We conclude that the injection occurs in a region close to the equator, not extending significantly towards the polar regions of the Be star;
    \item Injection radius: models with mass injection close to the stellar surface are a better match to the observed 1:1 correlation between the V/R oscillations in H$\alpha$ and the orbital frequency at the stellar equator. {Thus, mass injection at a distance of 0.25 \req or greater from the stellar surface does not reproduce the data, unless the equatorial radius of the sample in Paper I was systematically underestimated.} 
\end{enumerate}

The model that best matches the observational data of our reference, f\,Car, has a mass decretion rate of $\sim 6.9 \times 10^{-7} \, \rm M_\odot \, yr^{-1} \,\mathrm{str}^{-1}$, and $\gamma = 1.05$, namely material is injected with rotational velocities 1.05 times larger than the orbital speed at the injection radius, \rinj. Material circularises, i.e. the disk smooths out azimuthally due to the combined effect of orbital phase mixing and viscous diffusion, a couple of days after the end of mass ejection, at which point the material no longer holds information on its starting dynamics and distribution.

\subsection{Štefl frequencies}

Variations in emission asymmetry in H$\alpha$ and other emission lines are present in all the data shown in Paper~I. Some of the frequencies connected to these variations are known in the Be literature as Štefl frequencies \citep{stelf1998, baade2016}, and have always been interpreted as arising from orbiting circumstellar material. While historically detected in spectra, Štefl frequencies can also be seen in photometry, for instance in $\eta$ Cen and (possibly) $\alpha$ Eri \citep{baade2016}. These two stars are observed with high, near edge-on inclinations, meaning that an outburst in the equator could obscure part of the star. {Notably, the authors do not report photometric Štefl frequencies for $\mu$~Cen, which is not viewed edge-on.} {Štefl frequencies in spectroscopy were also detected in the sample studied in Paper I; Štefl frequencies in photometry could not be confirmed, although we note that the sample does not include edge-on stars with the exception of $\iota$\,Lyr. }

Variations of \ew in H$\alpha$ (thus frequencies we can assume are Štefl frequencies) are seen in all of our models (see Sect.~\ref{sec:results}), with the exception of the model with the largest opening angle, $\Delta\phi = 6.0$. Similarly, oscillations in photometry are also present in these models, particularly when seen at near edge-on inclination (see Fig.~\ref{fig:comp_big_85}). We propose that these photometric oscillations are associated with the concentrated ejecta partially eclipsing the star. Lomb-Scargle period analysis of the near edge-on light curves shown in the second row of Fig~\ref{fig:comp_big_85} recovers the same frequencies found in their respective \ew variations. These photometric frequencies are not detected for the light curves at lower inclinations. Therefore, we conclude that Štefl frequencies are indeed connected to material orbiting the star, and are more apparent in photometry for Be stars observed at high inclinations, where the magnitude of the partial eclipses is larger. 

\subsection{Caveats of the model}

The simplifications of our model (such as the velocity distribution of the ejecta, the isothermality of the disk, and the lack of a boundary layer between the photosphere and the disk) notwithstanding, there is a remarkable agreement between our preferred model and the observations. However, there are a couple of points of contention.
The first is that the shallower increase in EW after the end of mass injection seen in our simulations indicates that the simulated disk is dissipating slower than what we see in observations. Likely causes are higher viscosity than adopted in the models, radially variable viscosity, disk ablation, too intense or too long mass injection periods, and non-isothermal effects. 
{These physical processes, ablation in particular, may also explain the mismatch between the photometry (which is in good agreement with our template, f\,Car) and the too strong H$\alpha$ emission. Ablation would erode the disk, reducing the line emitting area. }

{The second point is the relation between the \ew variations and the orbital frequency. One of the conclusions of Paper~I is that the ejected material is initially in orbit around the star very close to its equator, since the frequencies of the \ew variations are, for all targets, very similar to the orbital frequency at the stellar equator \citep[$f_{\rm orb}$ --- see Fig. 10 of][]{labadie-bartz2025}. While the \ew frequency of our preferred model is a good match to $f_{\rm orb}$ (Sect.~\ref{sec:lever}), there is a phase shift in \ew right after the end of mass injection. Such a shift cannot be confirmed to exist in the observational data. This can be an effect of our assumption of an abrupt end to the mass ejection. Another possible cause is the inner boundary condition used in the SPH simulations, which assumes a torque-free interface. This is clearly an oversimplification, given the strong velocity gradient between the sub-Keplerian stellar surface and the Keplerian inner edge of the disk. A more physically realistic boundary condition could allow for coupling between the inner disk dynamics and the stellar rotation, potentially affecting the frequency and nature of the observed line profile asymmetries.}

As stated, our goal was to define the necessary conditions for mass to enter orbit such that the observations of Be outbursts were reproduced. We do not link these conditions to the physical cause of the Be phenomenon. After the end of mass ejection, the dynamics of the circumstellar material are governed by viscosity. The circularised material has no memory of the mechanism behind the mass ejection. As such, the results of our simulations hold for whatever mechanism triggered this process.

\subsection{The cause(s) of the Be phenomenon}

The correlation between the orbital frequency at the stellar equator and the \ew variation frequency reported in Paper~I indicates that Be stars, as previously expected from earlier works and observations, are fast but not critical rotators \citep[][]{fremat2005, rivinius2013, zorec2016}. If the Be star itself is not rotating critically, then a significant angular velocity must be added to the material in the injection volume; our models quantify this for the first time, suggesting that slightly super-Keplerian mass injection might be required, {plus a ballistic component of $20\, \rm km\, s^{-1}$}. With the injection volume so close to the equator, this velocity differential must be explained by a process that can provide this additional AM to the material. For instance, in the case of a $\rm 7.6\, M_{\odot}$ and $\rm 5.8 \, R_{\odot}$ star rotating with $v_{\rm rot}/v_{\rm orb} = 0.99$, an additional $\sim 50 \, \rm km\,s^{-1}$ is still required to send material from the equator into orbit {and build a sizeable disk}.

The two main mechanisms proposed to power the mass ejection of Be stars, in addition to their rapid rotation, are surface magnetic fields and NRPs. Large scale magnetic fields can be dismissed outright, as no such fields have ever been detected in Be stars, despite significant effort \citep[e.g.][]{wade2016}. Fields larger than 10\,G have been shown in MHD simulations to destroy, rather than form, a Keplerian decretion disk \citep{ud-doula2018}. \citet{balona2020} proposed that the photometric observations of Be outbursts by TESS can be explained by a model with two corotating clouds opposite one another, close to the stellar surface, held in place by a dipolar magnetic field \citep[first proposed in][]{balona2003}. However, such co-rotating structures would produce oscillations in peak separation that would be much more significant than the \ew oscillations, which are not observed in the data. \citet{nixon2020} proposed that a small scale field could inject AM rich material at a radius above the corotation radius of the star (similarly to solar coronal mass ejections), with the magnetic torque in the star/disk boundary continuously maintaining the material in orbit. When magnetic activity stops, the disk dissipates. However, no sort of solar corona-like ejections and flares has ever been observed in Be stars, {even now with high-cadence space photometry from TESS for nearly all known Galactic Be stars}. {Importantly, our lever-arm injection model fails to reproduce the observed \ew variation frequencies of the data, although this result is sensitive to the choices of stellar parameters and mass injection rate. While we cannot outright rule out small scale surface magnetic fields as playing some part in the mass ejection, there is currently no observational finding that strongly advocates for them. Improving our estimates of the fundamental parameters of the central star (the radius in particular) would allow us to confirm the correlation between stellar rotation and \ew variation frequencies, providing stronger constraints to the mass ejection mechanism.}

An NRP-powered mass ejection is an attractive scenario as it would neatly tie together the two ubiquitous characteristics of Be stars (fast rotation and NRPs) into their very \textit{raison d'être}. While linear NRP is likely insufficient to lift material into orbit, a non-linear combination of multiple modes might be able to achieve this \citep{baade2016, baade2018c}. There are now several observational examples of Be stars with clear connections between their pulsational frequencies and mass ejection events. For instance, $\mu$ Cen was the target of an in depth spectral analysis in \citet{rivinius1998a, rivinius1998b, rivinius2001}, where its line profile outbursts were found to be correlated with the beating of NRP modes with periods around 0.5\,d. $\eta$ Cen shows a similar multi-periodic behaviour \citep{rivinius2003}, which was also found to be connected to a combination of NRP modes \citep{baade2016}. More recently, \citet{richardson2021} found that one of the frequencies of the outbursts of HD\,6226 (a rare case of a Be star with nearly periodic outbursts) was equal to the difference of two pulsational frequencies seen in photometry. Baade et al. (in prep.) find that tens of pulsation modes are synchronised during Be outbursts. In their sample of 430 Be stars, \citet{labadie-bartz2022} found a correlation between the emergence and/or enhancement (both in amplitude and range of frequencies) of frequency groups in TESS data and outbursts. 
{However, the time baseline of the TESS observations was short, and individual closely-spaced frequencies could not be resolved. It is also possible that the observed emergence and/or enhancement in photometric signals could be a photospheric response to mass ejection.}
If such NRP modes have different wavelengths in the photosphere, their periodically developing synchronisation regions are also likely limited in azimuth, which could explain the localised ejections seen in the data in Paper~I, and supported by our simulations. 
Radial and horizontal displacements of surface layers of stars that show $g$ and $r$-modes are also expected to be affected by rotation \citep{saio2018a}, which is likely significant for Be stars.  

Modelling pulsations in Be stars is not trivial, but there have been theoretical developments in recent years. For instance, the photometric behaviour of HD\,49330 in CoRoT data \citep[][]{huat2009} was modelled by \citet{neiner2020}, following the methodology of \citet{neiner2012b}. They suggest that $g$ and $r$ waves stochastically excited in the core transport AM to the stellar envelope, spinning it up. The surface layers are destabilised by the accumulation of AM, leading to the emergence of transient $g$ modes. This combination causes the surface layers at the equator to reach critical velocity, leading to the outburst as a means of shedding the extra AM. That is, super-critical rotation is reached in the outer envelope during outburst, returning to sub-critical once enough AM and matter are shed \citep[this is also the case in the hypothesis of][]{saio2018b}. Transport of AM from stellar core to surface is crucial for the Be phenomenon, as repeating outbursts can only happen if AM is resupplied after each event. \citet{neiner2020}'s theory explains how AM can be deposited in the envelope on timescales of years: their model predicts an interval of around 11 years between outbursts for HD\,49330, not dissimilar to its observed interval of $\sim6$ years. However, these timescales are much larger than those of the flickers observed in Paper~I, indicating that other mechanisms are necessary to explain them.

{While there is no decisive observational or theoretical evidence of NRPs being the cause of the Be phenomenon, the synchronization of multiple NRP modes at the times of outbursts is a very strong indicator of a causal relation. However, the photometric variability of Be stars is very similar to that of Bn stars: rapidly rotating, pulsating B stars that are not Be stars. Frequency groups, for instance, which are correlated to outbursts in Be stars \citep{labadie-bartz2022}, are also present in Bn stars, but with no corresponding outburst {nor strong increases in frequency group strength beyond what can be explained by simple beating} \citep{naze2024}. Whether this indicates the pulsational properties of Be and Bn stars are different at a level that is difficult to discern \citep[e.g.][who suggested that only the Be stars pulsate in low-order NRP modes]{penrod1987}, or if another physical ingredient is necessary to make a B star into a Be is still unclear. Further work focusing on Be and Bn pulsations and asteroseismic modelling are needed to unequivocally establish if pulsation is the trigger mechanism of Be outbursts. }

{By bridging the gap between small-scale surface activity and large-scale disk dynamics, our} models provide much needed constraints on the geometry, dynamics, and mass ejection rate of flickers in Be stars, which {will serve as fundamental benchmarks for} future works that aim to connect the internal physical processes of these stars with their episodic outbursts. 

\begin{acknowledgements}
{The authors thank Dr. Peter Kroll for sending us his PhD thesis, which gave us a basis for the models developed here.} ACR acknowledges Miguel Caveagna Rubio for Fig~\ref{fig:esquema}.
This work made use of the computing facilities of the Centro de processamento de Dados do IAG/USP (CPD-IAG), whose purchase was made possible by the Brazilian agency FAPESP (grants 2019/25950-8,  2017/24954-4 and 2009/54006-4).
ACR acknowledges support from the `Coordenação de Aperfeiçoamento de Pessoal de Nível Superior' (CAPES grant 88887.464563/2019-00), the `Deutscher Akademischer Austauschdienst' (DAAD grant 57552338), from the studentship program at the European Southern Observatory, and from the Max Planck Institute for Astrophysics. {ACR acknowledges funding from the Netherlands Organisation for Scientific Research (NWO), as part of the Vidi and Aspasia research program BinWaves (project number 639.042.728, PI: de Mink). ACR acknowledges support from the Leverhulme Trust (grant No. RPG-2021-380)}.
ACC acknowledges support from CNPq (grant 314545/2023-9) and FAPESP (grants 2018/04055-8, 2019/26492-3, 2023/05083-3).  
JLB acknowledges support from the European Union (ERC, MAGNIFY, Project 101126182). 
Views and opinions expressed are, however, those of the authors only and do not necessarily reflect those of the European Union or the European Research Council. Neither the European Union nor the granting authority can be held responsible for them.
THA acknowledges support from FAPESP (grant 2021/01891-2).
Parts of the results in this work make use of the colourmaps in the CMasher \citep{cmasher2020} and Seaborn \citep{seaborn2021} packages.
\end{acknowledgements}

\bibliographystyle{aa}
\bibliography{bib}

\appendix

\FloatBarrier
\section{Definition of the isotropic velocity vector}\label{apd:def}

{The SPH particles are injected with an isotropic ballistic velocity in the co-rotating frame of the injection volume. This appendix describes how the propagation direction of this isotropic component is generated. The procedure guarantees that all directions are uniformly distributed over the outward hemisphere defined by the local stellar surface.}

To define the direction of propagation of the isotropic velocity vector, we first choose a random point at the inner rim of the injection volume, based on two random angles 
\begin{equation}
    \phi = \phi_0+\Delta\phi \xi_1\,,
\end{equation}
and
\begin{equation}
    \mu = \cos{\theta} = \frac{\Delta z}{R_{\rm inj}}\left(2 \, \xi_2  - 1\right) \,,
\end{equation}
\noindent
{where $R_{\rm inj}$ is the radius of the injection volume,} 
$\phi_0$ is the central $\phi$ of the injection volume, and $\xi_1$ and $\xi_2$ are two independent {random numbers uniformly distributed between 0 and 1.}. The injection volume rotates, so $\phi_0$ varies in time as
\begin{equation}\label{eq:time}
    \phi_0(t) = \phi_0(0) + \Omega_{\rm inj} \, t = \phi_0(0) + \gamma \frac{v_{\rm orb}}{R_{\rm inj}} \, t\,,
\end{equation}
where $v_{\rm orb} = (GM/R_{\rm inj})^{1/2}$.
{The factor $\gamma$ specifies the angular velocity of the injection volume relative to the local Keplerian orbital velocity.}
In Cartesian coordinates, for a particle $i$,  
\begin{equation}
    \begin{array}{lcr}
    x_i = R_{\rm inj} \sqrt{1 - \mu^2} \cos{\phi},\\
    y_i = R_{\rm inj} \sqrt{1 - \mu^2} \sin{\phi},\\
    z_i = R_{\rm inj} \mu\,.\\
\end{array}
\end{equation}
{A local tangent plane is then constructed at this point. Within this plane, two additional random angles, $\mu'$ and $\phi'$, define the direction of propagation of the isotropic velocity vector.}
With $a = \sqrt{1 - \mu^2}$ and $a' = \sqrt{1 - \mu'^2}$,

\begin{equation}
\hat{s} = 
\begin{cases}
  s_{x_i} = -a \sin{\phi}\sin{\phi'} + \cos{\phi}\left(a' \mu\cos{\phi'} + a \mu'\right)\,, \\
  s_{y_i} = a \cos{\phi}\sin{\phi'} + \sin{\phi}\left(a' \mu\cos{\phi'} + a \mu'\right)\,, \\
  s_{z_i} = \mu' \mu - a' a \cos{\phi'} \,.
  \end{cases}
\end{equation}
\noindent
{Since $\hat{s}$ is a unit vector, the isotropic velocity component in the stellar reference frame is simply}
\begin{equation}
\mathbf{v}_{\rm iso}=|v_i|\,\hat{s}.    
\end{equation}

{The initial particle velocity in the inertial frame is obtained by adding the orbital motion of the injection volume to this isotropic component}

\begin{equation}\label{eq:direc}
\begin{array}{lcl}
  v_x &=& |v_i| s_{x_i} - \gamma v_{\rm orb} \sin(\phi)\,, \\
  v_y &=& |v_i| s_{y_i} + \gamma v_{\rm orb} \cos(\phi)\,,\\
  v_z &=& |v_i| s_{z_i} \,.\\
  \end{array}
\end{equation}

{For convenience, the corresponding cylindrical velocity components are}

\begin{equation}
\begin{array}{lcl}
v_r &=& |v_i|\left(s_{x_i}\cos\phi+s_{y_i}\sin\phi\right)\,,\\
v_\phi &=& |v_i|\left(-s_{x_i}\sin\phi+s_{y_i}\cos\phi\right)
+\gamma v_{\rm orb}\,,\\
v_z &=& |v_i|\,s_{z_i}\,.
\end{array}
\label{eq:direc2}
\end{equation}

\FloatBarrier
\section{Additional figures}

Here we present additional figures that are helpful to illustrate points of our work, but not essential to the understanding of our methods, results and conclusions. 
The simulated light curves of our preferred model are compared directly to the TESS data for f\,Car in  Fig.~\ref{fig:flux_ess}. Figures \ref{fig:besthdust_zoom},  \ref{fig:comp_big_85}, and \ref{fig:tall} are complements of Figs. \ref{fig:besthdust}, \ref{fig:comp_big}, and \ref{fig:bestsph}, as described in their captions.



\begin{figure}
    \centering
    \resizebox{\hsize}{!}{\includegraphics{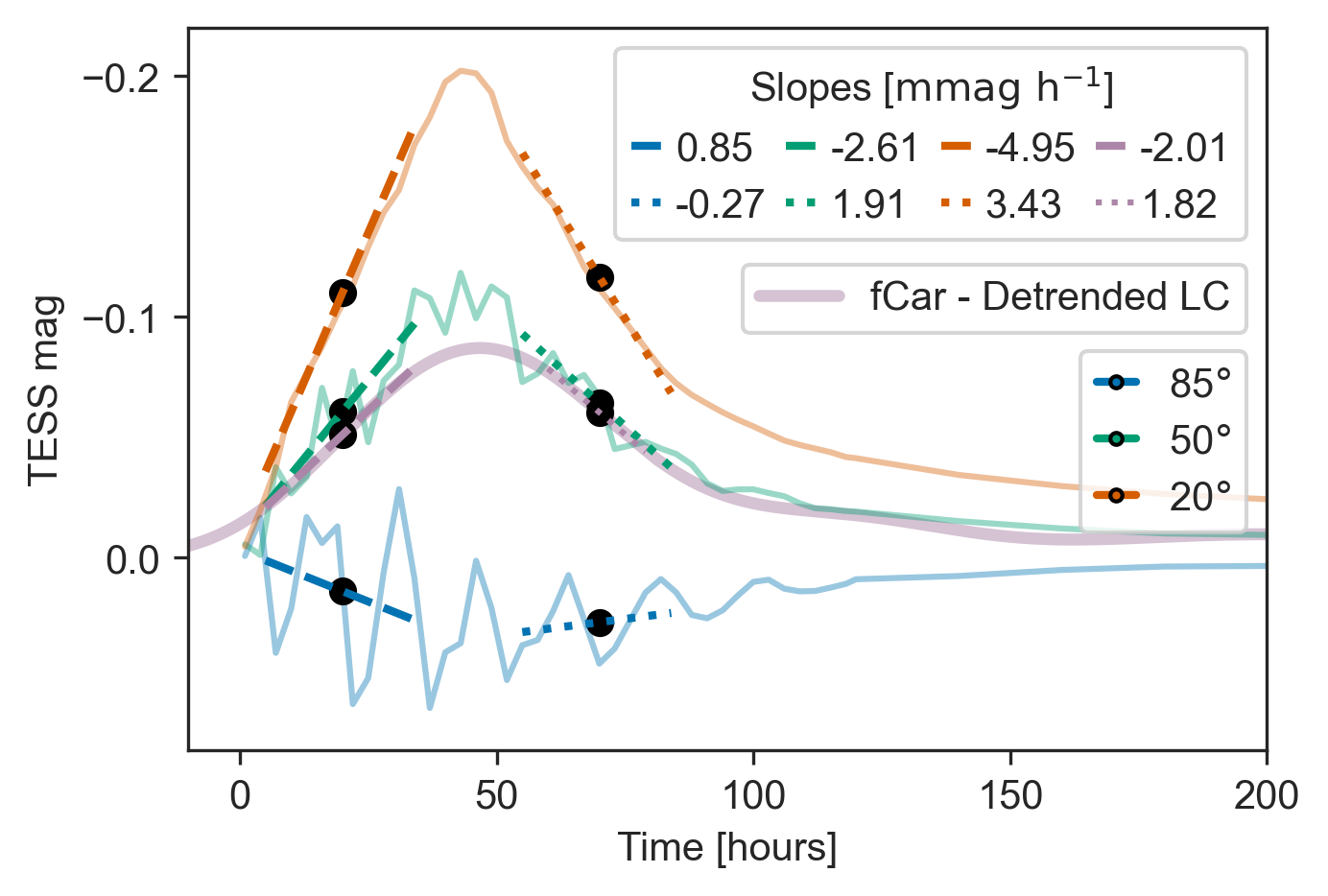}}
    \caption{Simulated TESS light curve for our preferred model, described in Sect.~\ref{sec:most_fave}, for three inclination angles (20, 50 and 85\degree). We show the measured slopes of build-up and dissipation (at 20 and 70 h), and also show the f\,Car flicker, in grey, as a reference.}
    \label{fig:mags_slopes}
\end{figure}

\begin{figure}
    \centering
    \resizebox{\hsize}{!}{\includegraphics{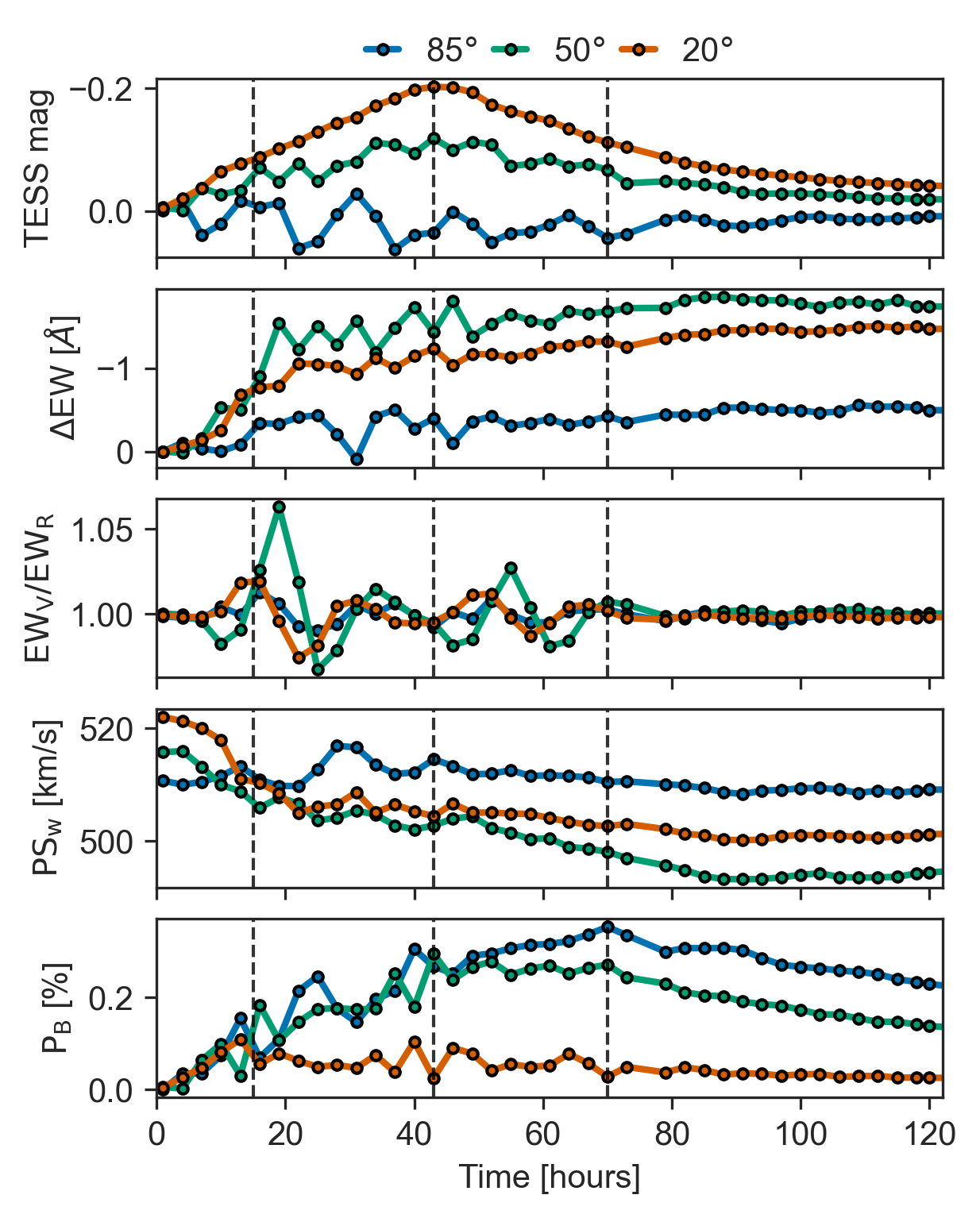}}
    \caption{Same as the left part of Fig.~\ref{fig:besthdust}, but zoomed in to the first 120\,h of the simulation.}
    \label{fig:besthdust_zoom}
\end{figure}



\begin{figure*}
    \centering
    \resizebox{\hsize}{!}{\includegraphics{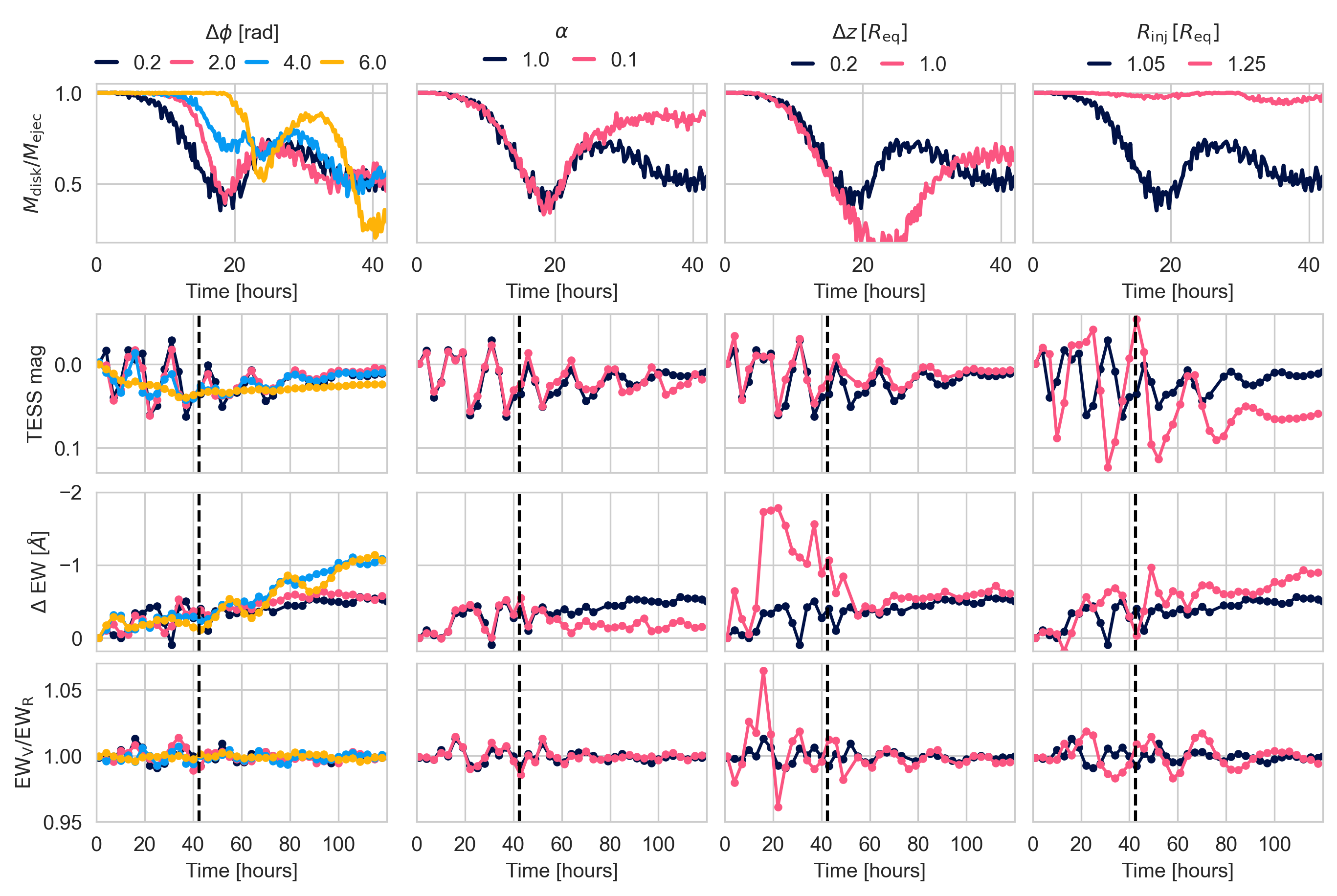}}
    \caption{Same as Fig.~\ref{fig:comp_big}, but for the inclination angle of 85\degree.}
    \label{fig:comp_big_85}
\end{figure*}

\begin{figure*}
    \centering
    \resizebox{\hsize}{!}{\includegraphics{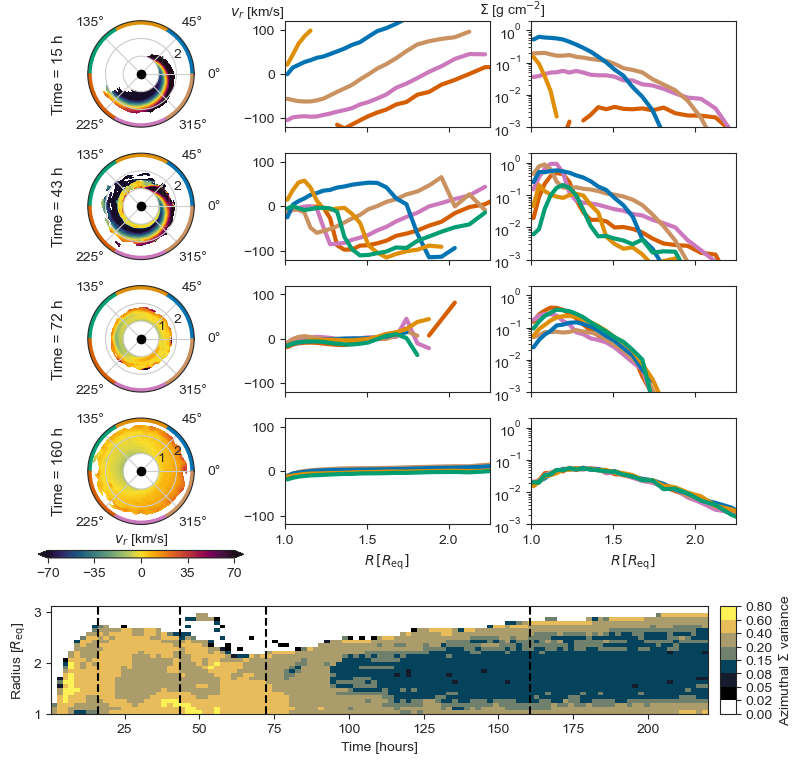}}
    \caption{Same as Fig.~\ref{fig:bestsph}, but for a model with $\Delta z = 1.0$\req. The simulation evolves similarly to our preferred model, with $\Delta z = 0.2$\req.}
    \label{fig:tall}
\end{figure*}

\FloatBarrier
\vfill\clearpage

\section{Disk temperatures and emitting areas}

\begin{figure}
    \centering
    \resizebox{\hsize}{!}{\includegraphics{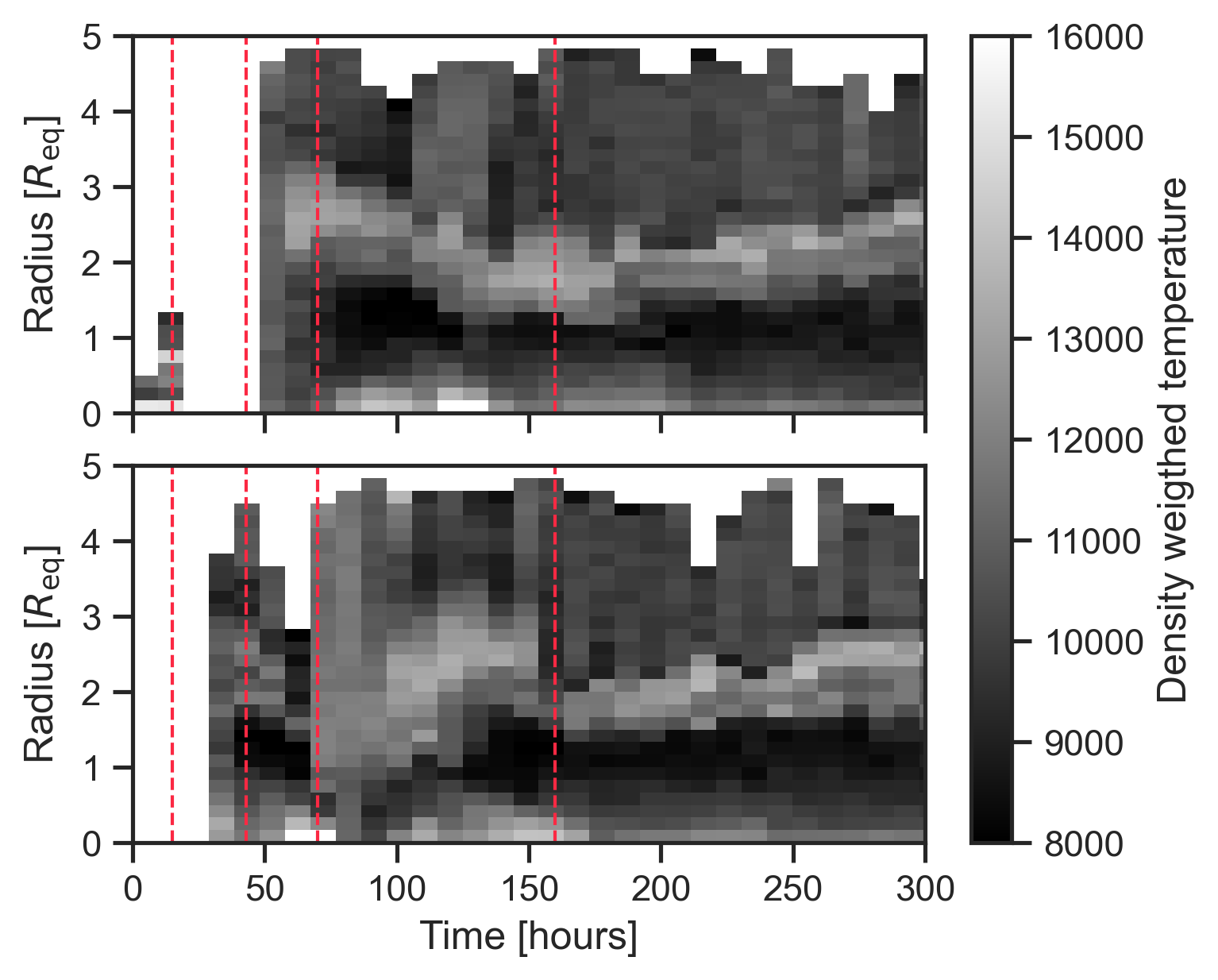}}    \caption[Maps of the density weighted temperature of the disk.]{Maps of the density weighted temperature of the disk for the preferred simulation. The top panel shows the average temperature calculated inside a wedge with an opening angle of 25$\degree$, centred on the injection volume. It rotates along with the injection volume following Eq.~\ref{eq:time}. The bottom panel shows the same, but for a wedge pointed in the opposite direction to the injection volume, i.e. with an 180$\degree$ difference in azimuth. The red dashed lines mark the four snapshots highlighted in Fig.~\ref{fig:bestsph}.}
    \label{fig:temp}
\end{figure}

The SPH simulations presented here are isothermal. However, to create the synthetic observables, HDUST calculates a 3D temperature profile in the material. Figure~\ref{fig:temp} shows maps of the density-weighted temperature calculated inside a wedge with an opening angle of 25$\degree$ {in azimuth}, centred on the injection volume that rotates along with it (top panel), and a wedge pointed in the opposite direction to the injection volume (bottom panel). This illustrates a caveat of our SPH models, as the temperature structure of the disk is far from isothermal and would therefore lead to density and velocity variations that are not being considered in the hydrodynamic calculations. Figure~\ref{fig:temp} also showcases how the circularisation of the material includes a thermal as well as a dynamical homogenization.

Along with the temperature, the density and emitting area of the material control the photometric amplitude of a flicker; this means that the amplitude depends on the size of the area with optical depth $\tau \gtrsim 1$ for a given wavelength. We calculated the optical depth using hydrogen level populations from HDUST and a custom ray-tracing code \citep{deamorim2025} that accounts for free-free, bound-free, and bound-bound absorption and electron scattering. Adopting the stellar parameters for f\,Car (Sect.~\ref{sec:model_descrip}) and an inclination of 50$\degree$, we computed the optical depth along rays that traverse the disk from the far side towards the observer. For rays that intersect the central star, the optical depth is integrated starting at the stellar surface. We focused on two diagnostic wavelengths: the $V$ band (540~nm, the Johnson $V$ central wavelength) and H$\alpha$. The latter is more complex due to the high velocities within the disk, which Doppler-shift the absorption profile. To capture a representative H$\alpha$ absorption optical depth, we considered the maximum optical depth within a velocity range of $\pm 500~ \rm km\,s^{-1}$ centred on the H$\alpha$ rest wavelength.

\begin{figure}
    \centering
    \resizebox{\hsize}{!}{\includegraphics{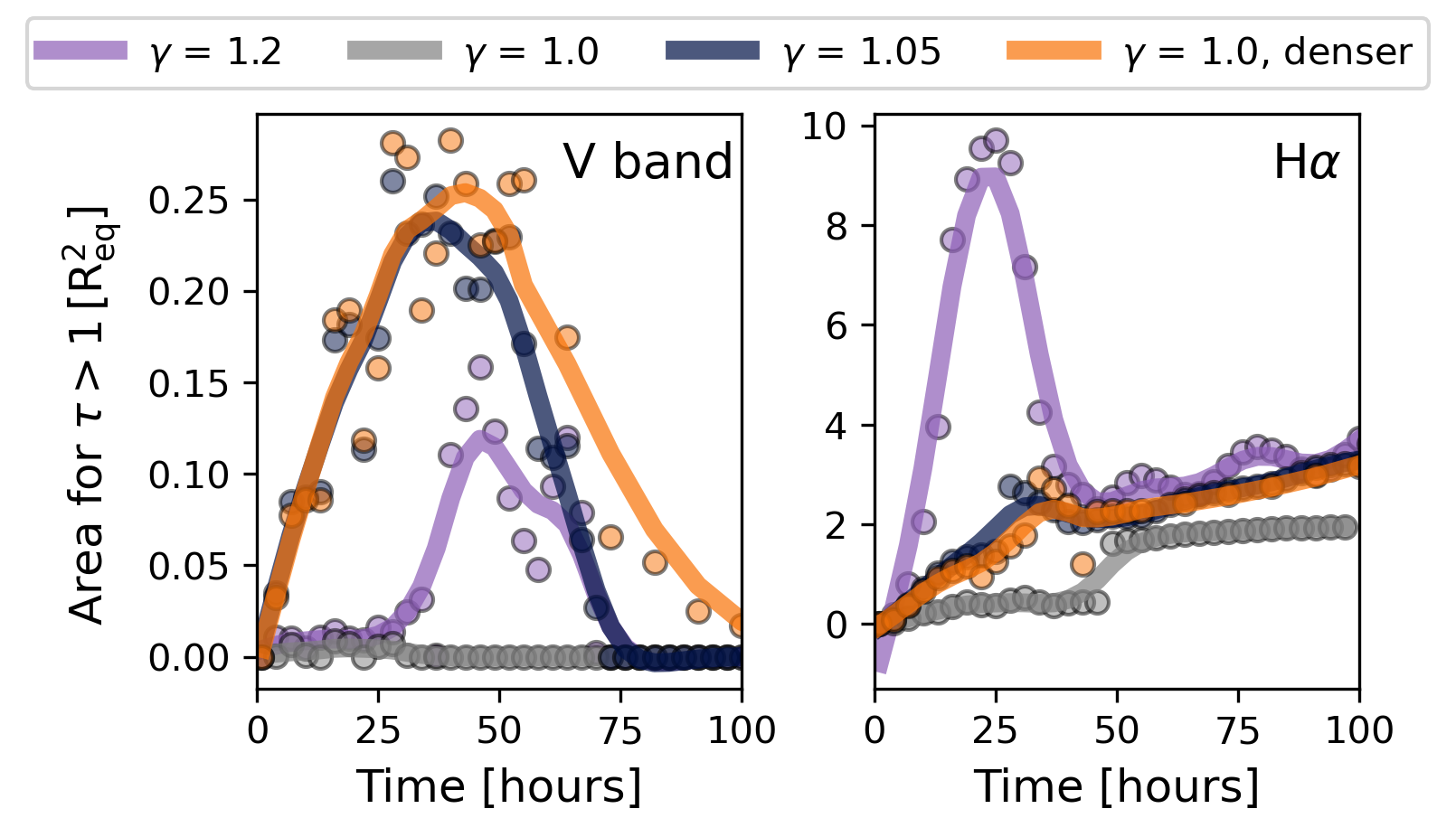}}
    \caption{Areas of the optically thick ($\tau>1$) regions in the $V$ band and at H$\alpha$ for simulations with $\gamma = 1.0$, 1.05, and 1.2, with fixed $\alpha = 1.0$\, \rinj = 1.05\req, $\Delta r = 0.01$\req, and $\Delta \phi = 0.2$ rad. The thick lines show a polynomial fit to the data points for ease of comparison between models. Models in orange and grey are based on the same SPH simulation (with $\gamma = 1.0$), but the orange model was made 9 times denser in the HDUST radiative transfer calculations, hence its larger optically thick area.
    }
    \label{fig:taucompare}
\end{figure}

The observables are sensitive to the effective mass injection rate adopted in the simulation, as the comparison between the two models with $\gamma = 1.0$ highlights. Comparing the shapes of the curves in Fig.~\ref{fig:taucompare} for the $V$ band and H$\alpha$ emitting areas with the photometric and EW curves of Fig.~\ref{fig:comp_small} for the same models, we find a very close match. Therefore, the sizes of the optically thick areas can be used as diagnostic tools to estimate the emission of regions of interest.


%
%

\end{document}